 \documentstyle[12pt]{article}
 \newcommand{\beqn}{\begin{eqnarray}}
 \newcommand{\eeqn}{\end{eqnarray}}
 \newcommand{\be}{\begin{equation}}
 \newcommand{\ee}{\end{equation}}
 \newcommand{\ba}{\begin{array}}
 \newcommand{\ea}{\end{array}}
 \newcommand{\pa}{\partial}
 \newcommand{\re}{\ref}
 \newcommand{\ci}{\cite}
 \newcommand{\la}{\label}
 \newcommand{\bfr}{\begin{flushright}}
  \newcommand{\efr}{\end{flushright}}
 \newcommand{\bfl}{\begin{flushleft}}
 \newcommand{\efl}{\end{flushleft}}
 \newcommand{\fr}{\frac}
 \newcommand{\ov}{\overline}
 \newcommand{\ve}{\varepsilon}
\newcommand{\eps}{\epsilon}
 \newcommand{\ga}{\gamma}
\newcommand{\de}{\delta}
 \newcommand{\al}{\alpha}
 \newcommand{\si}{\sigma}
\newcommand{\Si}{\Sigma}
 \newcommand{\ds}{\displaystyle}
 \newcommand{\pr}{\prime}
 \newcommand{\De}{\Delta}
\newcommand{\na}{\nabla}
\newcommand{\La}{\Lambda}
 
 \newcommand{\Om}{\Omega}
\newcommand{\om}{\omega}
\newcommand{\tr}{\mbox{t\hspace{-0.3mm}r}\hspace{0.3mm}}

\newcommand{\brr}{{|\kern-.15em|\kern-.15em|\kern-.15em}\,}

 \newcommand{\br}{|\kern-.25em|\kern-.25em|}
 
      \def\N{{\rm I\kern-.1567em N}}                              
 \def\R{{\rm I\kern-.1567em R}}                              
 \def\C{{\rm C\kern-4.7pt                                    
 \vrule height 7.7pt width 0.4pt depth -0.5pt \phantom {.}}
\hspace{0.4mm}}
 \def\Z{{\sf Z\kern-4.5pt Z}}                                

\newcommand{\supp}{\mathop{\rm supp}\nolimits}

 \newtheorem{theorem}{Theorem}[section]
 
 \newtheorem{definition}[theorem]{Definition}
 
 \newtheorem{lemma}[theorem]{Lemma}
 
 \newtheorem{remark}[theorem]{Remark}
 
 \newtheorem{cor}[theorem]{Corollary}
 \newtheorem{pro}[theorem]{Proposition}

\begin{document}
 \begin{titlepage}
\hspace{6cm} {\em Comm. Math. Phys.} 
{\bf 225} (2002), no.1, 1-32
\bigskip\bigskip\bigskip
\vspace{2cm}

{\it Dedicated to M.I.~Vishik on the occasion of his  $80$'s anniversary}
\vspace{0mm}

\begin{center}
 {\Large\bf  On Convergence to Equilibrium Distribution, I.
\medskip\\
The Klein -- Gordon Equation with Mixing}
  \end{center}
\vspace{0mm}
{\bf T.V.~Dudnikova
 \footnote{Supported partly by research grants of
  DFG (no.436 RUS 113/505/1) and
 RFBR (no.01-01-04002).}
}
Mathematics Department,
 Elektrostal Polytechnical Institute,\\
        Elektrostal, 144000 Russia;
e-mail:~misis@elsite.ru.
 \medskip\\
 {\bf A.I.~Komech
 \footnote{Supported partly by the
Institute of Physics and Mathematics of Michoacan in Morelia,
by Max-Planck Institute for the Mathematics in the Sciences
in Leipzig and by
research grant of DFG (no.436 RUS 113/505/1).}
 }
 Mechanics and Mathematics Department, Moscow State University,
 Moscow, 119899 Russia;
 e-mail:~komech@mech.math.msu.ru.
 \medskip\\
 {\bf E.A.~Kopylova
\footnote{Supported partly by research grant of
 RFBR (no.01-01-04002).}
 }
 Physics and Applied Mathematics Department,
Vladimir State University,
 Vladimir, Russia;
 e-mail:~ks@vpti.vladimir.ru.
\medskip\\
 {\bf Yu.M.~Suhov
 }
 Statistical Laboratory,
Department of Pure Mathematics and Mathematical Statistics,
University of Cambridge,
 Cambridge, UK;
 e-mail:~Y.M.Suhov@statslab.cam.ac.uk.
\vspace{12.5mm}
\begin{abstract}
Consider the Klein-Gordon equation (KGE)
in $\R^n$, $n\ge 2$, with  constant or variable coefficients.
We study the distribution $\mu_t$
 of the {\it random solution} at  time $t\in\R$.
We assume that the initial probability measure $\mu_0$
has zero mean, a translation-invariant covariance,
and a finite mean energy density.
We also asume that $\mu_0$ satisfies a Rosenblatt- or
Ibragimov--Linnik-type mixing condition.
The main result is the convergence of $\mu_t$ to
a Gaussian probability measure as $t\to\infty$
which gives a Central Limit Theorem for the KGE.
The proof for the case of constant coefficients
is based on an analysis of long time asymptotics
of the solution in the Fourier representation and
Bernstein's `room-corridor' argument. The case
of variable coefficients is treated
by using an `averaged' version of
the scattering theory for infinite
energy solutions, based on Vainberg's results on local energy decay.

 \end{abstract}
\end{titlepage}

 \section{Introduction}
The aim of this paper is
to underline a special role of equilibrium distributions
in statistical mechanics  of systems
governed by hyperbolic partial differential equations
(for parabolic equations see \ci{Bu,RSS}).
Important examples arise when one discusses
the role a {\it canonical} Gibbs distribution (CGD) in
the Planck theory of
spectral density of the black-body emission and  in the
Einstein--Debye quantum theory
of solid state (see, e.g. \ci{Seitz}). [The word `canonical'
is used in this paper to emphasize the fact that the
probability distribution under consideration
is formally related to the `Hamiltonian', or
the energy functional, of the corresponding
equation by the Gibbs exponential formula.
Owing to the linearity of our equations,
there are plenty of other
first integrals which lead to other stationary measures.]
Historically, the emission law was established
at a heuristical level
by Kirchhoff in 1859 (see \ci{Som})
and stated formally by Planck in 1900 (see \ci{Pl}).
The law concerns the
correspondence between the temperature
and the colour of an
emitting body (e.g., a burning carbon,
or an incandescent wire in an electric bulb).
Furthermore, it provides a fundamental
information on an interaction
between the Maxwell field and a `matter'.
Planck's formula
specifies a `radiation intensity' $I_T(\om)$ of the electromagnetic
field at a fixed temperature $T>0$,
as a function of the  frequency $\om>0$.
It is convenient to treat $I_T(\cdot)$
as the  spectral correlation function of
a stationary random process. Then
if  $g_T$  denotes an equilibrium distribution of this
process, the Kirchhoff-Planck law
suggests the long-time convergence
\be\la{1.1}\mu_t \rightharpoondown
g_T,\,\,\,\, t\to \infty.\ee
Here  $\mu_t$ is the distribution at time $t$ of a
nonstationary random solution. The resulting
equilibrium temperature $T$
is determined by an initial distribution $\mu_0$.
Convergence to equilibrium (\re{1.1}) is also expected
in a system of Maxwell's equations coupled to
an equation of evolution of a `matter'.
For example, both (\re{1.1}) and
the Kirchhoff--Planck law should hold for
the coupled  Maxwell--Dirac equations \ci{Bour1},
or for their second-quantised modifications.
However, the rigorous proof here is still an open problem.

Previously, the convergence of type (\re{1.1})
to a CGD $g_T$ has been established for an ideal gas
with infinitely many particles by
Sinai (see, e.g., \ci{CFS}).
Similar results were later obtained for
other infinite-dimensional systems
(see \ci{BDS,DS} and a survey \ci{DSS}).
For nonlinear wave problems, the first result of such kind
has been established by Jaksic and Pillet in \ci{JP}:
they consider a system of a classical particle
coupled to a wave field in a smooth nonlocal fashion.
For all these models, the CGD $g_T$  is well-defined,
although the convergence is highly non-trivial. On the
other hand, for the local coupling
such as in the Maxwell--Dirac equations, the problem of
`ultraviolet divergence' arises: the CGDs cannot be
defined directly as the local energy is formally infinite almost surely.
This is a serious technical difficulty that
suggests that, to begin with, one should analyse convergence to
{\it non-canonical} stationary measures $\mu_\infty$,
with finite mean local energy:
\be\la{1.2}\mu_t \rightharpoondown
\mu_\infty,\,\,\,\, t\to \infty.\ee
In fact, most of the above-mentioned papers establish the
convergence to both CGDs and non-canonical stationary
measures, by using the same methods. In our situation,
the aforementioned ultraviolet divergence makes the difference between
(\re{1.1}) and (\re{1.2}).

In this paper we prove convergence (\re{1.2})
for the Klein--Gordon equation (KGE) in $\R^n$, $n\ge 2$:
\beqn\la{1.3}\left\{\ba{l}
\ddot u(x,\,t)  =  \sum_{j=1}^{n}(\pa_j - iA_j(x))^2 u(x,t)
 - m^2\, u(x,t),~~ x \in\R^n,\\
u|_{t=0} = u_0(x),~~\dot u|_{t=0} = v_0(x).
\ea\right.\eeqn
Here $\ds\pa_j\equiv \fr\pa{\pa x_j}$, $x\in\R^n$, $t\in\R$,
$m>0$ is a fixed constant and $(A_1(x),\dots,A_n(x))$
a vector potential of an external magnetic field;
we assume that functions $A_j(x)$ vanish outside a
bounded domain. The solution $u(x,t)$ is considered
as a complex-valued {\it classical} function.

It is important to identify a natural property
of the initial measure $\mu_0$
guaranteeing convergence (\re{1.2}).
We follow an idea of Dobrushin and Suhov \ci{DS} and
use a `space'-mixing condition
of Rosenblatt- or Ibragimov--Linnik-type.
Such a condition is natural
from physical point of view. It replaces a `quasiergodic
hypothesis' and allows us to avoid introducing
a `thermostat' with a prescribed time-behaviour.
Similar conditions have been used in \ci{BDS, BPT, SL, SS}.
In this paper, mixing is defined and applied in the context
of the KGE.

Thus we prove convergence (\re{1.2})
for a class of initial measures
$\mu_0$ on a classical function space,
with a finite mean local energy and
satisfying a mixing condition.
The limiting measure $\mu_\infty$ is stationary and
turns out to be a Gaussian probability measure (GPM).
Hence, this result is a form of the Central Limit
Theorem for the KGE.

Another important question we discuss below is the relation
of the limiting measure $\mu_\infty$ to the CGD $g_T$.
The (formal) Klein--Gordon Hamiltonian is
given by a quadratic form and so the CGDs $g_T$
are also GPMs, albeit generalised
(i.e. living in generalised function spaces).
As our limiting measures $\mu_\infty$
are `classical' GPMs, they do not include
CGDs. However, in the case of constant coefficients,
a CGD can be obtained as a limit of measures $\mu_\infty$ as
the `correlation radius' figuring in the mixing
conditions imposed on $\mu_0$ tends to zero.
More precisely, we assume that for a fixed $T>0$
\be\la{1.4}
\fr 12 E\Big(v_0(x) v_0(y)+\na u_0(x)\cdot\na u_0(y)+
m^2u_0(x) u_0(y)\Big)\to
T\de(x-y),\,\,\,\,r\to 0,\ee
where $E$ denotes the expectation. Then the covariance
functions (CFs) of corresponding limit GPM $\mu_\infty$
converge to the covariance functions of the
CGD $g_T$. In turn, this implies the convergence
\be\la{1.5}\mu_t \rightharpoondown
\mu_\infty\sim g_T,\,\,\,r\to 1.\ee
See Section 4.

It should be noted that the existence of a `massive'
(in a sense, infinite-dimensional) set of the limiting
measures $\mu_\infty$ that are different from CGD's is
related to the fact that KGE (\re{1.3}) is degenerate
and admits infinitely many `additive' first integrals. Like
the Klein--Gordon Hamiltonian, these integrals are quadratic
forms; hence they generate GPMs via Gibbs exponential formulas.

Convergence (\re{1.2}) has been obtained  in \ci{KM,Ko,diss}
for translation-invariant initial measures $\mu_0$.
However, the original proofs were too long and used
a specific apparatus of Bessel's functions applicable
exclusively in the case of the KGE. They
have not been published in detail because of lack of a unifying argument
that could show the limits of the method and its
forthcoming developments. To clarify the mechanism
behind the results, one needed some new and robust ideas.
The current work provides a modern approach
applicable to a wide class of
linear hyperbolic  equations with a nondegenerate
`dispersion relation', see Eqn (\re{Hess}) below.
We also weaken considerably the mixing condition on measure $\mu_0$.
Moreover, our approach yields much shorter proofs and
is applicable to non-translation invariant initial measures.
The last fact is important in  relation to
two-temperature problem \ci{BPT,DKS,SL}
and hydrodynamic limit \ci{DPS}. Such progress became possible
in a large part owing to a systematic use of
a Fourier transform (FT) and a duality argument of Lemma \re{ldu}.
[The importance of the Fourier transform
were demonstrated in earlier works \ci{BPT,SS,SL}.]
\medskip

Similar results, for the wave equation (WE) in $\R^n$
with odd $n\ge 3$, are established in  \ci{DKRS}
which develops the results \ci{R1}.
The KGE shares some common features with the WE
(which is formally obtained by setting $m=0$ in
(\re{1.3})), and
the exposition in \ci{Ko,diss} followed the structure
of the earlier work \ci{R1}. On the other hand,
the KGE and WE also have serious differences, see below.

It is worth mentioning that possible extensions of our methods
include, on
the one hand, Dirac's and other relativistic-invariant
linear hyperbolic equations
and on the other hand harmonic lattices, as well as `coupled' systems
of both types. We intend to return to this problems elsewhere.
\medskip

We now pass to a detailed description of results.
Formal definitions and statements are given in Section 2. Set:
$Y(t)=(Y^0(t),Y^1(t))\equiv (u(\cdot,t),\dot u(\cdot,t))$,
$Y_0=(Y^0_0,Y^1_0)\equiv (u_0,v_0)$.
Then (\ref{1.3}) takes the  form of an evolution equation
\be\la{CP}
\dot Y(t)={\cal A}Y(t),\,\,\,t\in\R;\,\,\,\,Y(0)=Y_0.\ee
Here,
\be\la{A}{\cal A}=\left(
\begin{array}{cc}
0 & 1\\
A & 0\end{array}\right),\ee
where $A=\sum_{j=1}^{n}(\pa_j - iA_j(x))^2-m^2$.
We assume that the initial date $Y_0$ is a
random element of a complex functional space ${\cal H}$
corresponding to states with a finite local energy,
see Definition \ref{d1.1} below.
The distribution of $Y_0$ is a probability measure  $\mu_0$
of mean zero satisfying some additional assumptions, see
Conditions {\bf S1}-{\bf S3} below.
Given $t\in\R$, denote by $\mu_t$ the measure
that gives the distribution of $Y(t)$, the
random solution to  (\re{CP}). We study the
asymptotics of $\mu_t$ as $t\to\pm\infty$.

We identify $\C\equiv \R^2$ and denote by $\otimes$ tensor
product of real vectors. The CFs of the initial measure
are supposed to be translation-invariant:
\be\la{1.9'}
Q^{ij}_0(x,y):= E\Big(Y_0^i(x)\otimes
{Y_0^j(y)} \Big)=
q^{ij}_0(x-y),\,\,\,x,y\in\R^n,\;\;i,j=0,1,\ee
(in fact our methods require a weaker assumption,
but to simplify the exposition, we will not discuss
it here). We also assume that the
initial mean energy density  is finite:
\be\la{med}
e_0:=E \Big(
\vert v_0(x)\vert^2+\vert \nabla u_0(x)\vert^2+
m^2\vert u_0(x)\vert^2
 \Big)=q_0^{11}(0)
-\De q_0^{00}(0)+m^2q_0^{00}(0)
<\infty,\,\,\,\,x\in\R^n.
\ee
Finally, we assume that measure $\mu_0$ satisfies a
mixing condition of a Rosenblatt- or Ibragimov-Linnik type, which means that
\be\la{mix}Y_0(x)\,\,\,\,   and \, \, \,\,Y_0(y)
\,\,\,\,  are\,\,\,\, asymptotically\,\,\,\, independent\,\, \,\,
 as \,\, \,\,|x-y|\to\infty.\ee
As was said before, our main result gives the (weak) convergence
(\re{1.2}) of $\mu_t$ to a limiting measure $\mu_\infty$
which is a stationary GPM on
${\cal H}$. A similar convergence holds for $t\to-\infty$.
An explicit formulas are then given for the CFs of $\mu_\infty$.
\medskip

The strategy of the proof is as follows.
First, we prove (\re{1.2}) for the equation with
constant coefficients ($A_{k}(x)\equiv 0$), in three steps.
\\
{\bf ~~I.}
We check that the family of measures $\mu_t$, $t\geq 0$,
is weakly compact.\\
{\bf ~II.}
We check that the CFs  converge to a limit:
for $i,j=0,1$,
\be\la{corf}Q_t^{ij}(x,y)=\int Y^i(x)\otimes
{Y^j(y)}\mu_t(dY)
\to Q_\infty^{ij}(x,y),\,\,\,\,t\to\infty.\ee
{\bf III.} Finally, we check that the
characteristic functionals converge to a Gaussian one:
\be\la{2.6i}
\hat\mu_t(\Psi ):=
\int \exp\{i\langle Y,\Psi\rangle \}\mu_t(dY)
\rightarrow \exp\{-\fr12 {\cal Q}_\infty( \Psi, \Psi)\},
\,\,\,t\to\infty. \ee
Here $\Psi$ is an arbitrary element of the dual space and
${\cal Q}_\infty$ the quadratic form
with the integral kernel
$(Q^{ij}_\infty(x,y))_{i,j=0,1}$;
$\langle Y,\Psi \rangle$ denotes scalar product
in a real Hilbert space $L^2(\R^n)\otimes\R^N$.
\medskip

Property {\bf I} follows from the Prokhorov Theorem
by a method  used in \ci{VF}. First, we prove
a uniform bound for the mean local energy in $\mu_t$, using
the conservation of mean energy density.
The conditions of the Prokhorov Theorem are then checked by
using Sobolev's embedding Theorem in conjunction with
Chebyshev's inequality. Next, we deduce property {\bf II}
from an analysis of oscillatory integrals arising in the FT.
An important role is attributed to Proposition \re{l4.1}
reflecting the properties of the CFs in the FT
deduced from the mixing condition.

On the other hand, the FT approach alone
is not sufficient for proving property {\bf III}
even in the case of constant coefficients.
The reason is that
a function of infinite energy corresponds to a singular
generalised function
in the FT, and the exact interpretation
of mixing condition
(\re{mix}) for such generalised  functions is unclear.
We deduce property {\bf III}
from a representation of the solution
in terms of the  initial date
in the coordinate space. This  is a modification of
the approach adopted in \ci{KM, Ko,diss}.
It allows us to combine the
mixing condition with the fact that waves
in the coordinate space disperse to infinity.
This leads to a representation of the solution
as a sum of weakly dependent random variables.
Then (\ref{2.6i}) follows from a Central Limit Theorem (CLT)
under a Lindeberg-type condition.
Checking such a condition is an important part
of the proof.

It is useful to
discuss the dispersive mechanism that is behind (\ref{2.6i})
and compare the KGE ($m>0$) and WE ($m=0$).
Take, for  simplicity, $n=3$ and
$u_0\equiv 0$. The solution to (\ref{1.3})
(with $A_k(x)\equiv 0$)
is given by
\be\la{E}
u(x,t)=\int{\cal E}(x-y,t)v_0(y)\,dy,\quad t>0,
\ee
where ${\cal E}$ is the `retarded'
fundamental solution
\be\la{fs}
{\cal E}(x,t)=
\frac{1}{4\pi t}\delta(|x|-t)
-\frac{m\theta(t-|x|)}{4\pi}
\frac{J_1(m\sqrt{t^2-x^2})}{\sqrt{t^2-x^2}},
\ee
$J_1$ is the Bessel function of the first order.
For $m=0$ function  ${\cal E}(\cdot,t)$
is supported by the sphere $|x|=t$ of area $\sim t^2$,
and (\ref{E}) becomes
the Kirchhoff formula
\be\la{K}
u(x,t)=\frac{1}{4\pi t}\int\limits_{|x-y|=t}v_0(y)\,dS(y),
\ee
which manifests the dispersion of waves in the 3D space.
Dividing the sphere $\{y\in\R^3:~|x-y|=t\}$
into $N\sim t^2$
`rooms' of a fixed width $d\gg 1$,
 we rewrite (\ref{K}) as
\be\la{CLT}
u(x,t)\sim\frac{\sum\limits_{k=1}^{N}r_k}{\sqrt{N}},
\ee
where $r_k$ are nearly independent owing to the mixing condition.
Then (\re{1.2}) follows by
well-known
Bernstein's `room-corridor' arguments.

For $m>0$ function
 ${\cal E}(\cdot,t)$
is supported by the ball $|x|\le t$
 which  means the absence of a
 {\it strong} Huyghen's principle for
the KGE. The volume of the ball is $\sim t^3$, hence
rewriting (\ref{E}) in the
form (\ref{CLT}) would need asymptotics of the type
\be\la{**}
{\cal E}(x,t)={\cal O}(t^{-3/2}),\quad |x|\le t
\ee
as $t\to\infty$.
As $J_1(r)\sim\cos(r-3\pi/4)/\sqrt{r}$,
asymptotics (\re{**}) only holds in the region
$|x|\le vt$ with $v<1$. For instance,
$$
{\cal E}\bigr|_{|x|=vt}\sim
\frac{\cos(m\gamma t-3\pi/4)}{(\gamma t)^{3/2}},
$$
where  $\gamma=\sqrt{1-v^2}$.
However, the degree of the decay is different
 near the light cone $|x|=t$
corresponding to
$v=1$
and $\gamma=0$. For example,
for a fixed $r>0$,
\be\la{OS}
{\cal E}\bigr|_{|x|=t-r}\sim
\frac{\cos(m\sqrt{2rt}-3\pi/4)}{(2rt)^{3/4}}={\cal O}(t^{-3/4}),
\ee
where $r=t-|x|$ is the `distance'
from the light cone.
This illustrates that an application of
Bernstein's method
in the case of the KGE
requires a new  idea.

The key observation is
that the  asymptotics (\ref{OS}) displays  oscillations
$\sim \cos m\sqrt{2rt}$
of ${\cal E}$ near light cone as $t\to\infty$.
The solution becomes an oscillatory integral, and one is able
 to
compensate the weak decay $\sim t^{-3/4}$ by a partial
integration with Bessel functions, by method
following an argument from \ci[Appendix B]{MS}.
Such an  approach was used in \ci{diss} and was
accompanied by tedious
computations in a combined `coordinate-momentum'
representation.
The approach adopted in this paper
 allows us
  to avoid
this part of the argument. An important role plays
 a duality argument
of Lemma \re{ldu} leading to
an analysis of an oscillatory integral
with a phase function (=`dispersion relation')
with a
nondegenerate Hessian, see (\re{Hess}).

Simple examples show that  the convergence  may fail when
 the mixing condition does not hold.
For instance,
take $u_0(x)\equiv \pm 1$ and
$v_0(x)\equiv 0$ with probability $p_\pm=0.5$
then  mean value is zero and (\re{med}) holds, but (\re{mix}) does not.
The solution
$u(x,\,t)\equiv\pm \cos~(mt)$ a.s.,
hence $\mu_t$ is periodic in time, and (\re{1.2}) fails.
\medskip

Finally, a comment on the case
of variable coefficients $A_k(x)$.
In this case explicit formulas for the solution
are unavailable.
Here
we construct
a scattering theory for solutions of
infinite global
energy.
This version of the  scattering theory allows us to reduce
the proof of (\re{1.2}) to the case
of constant coefficients
(this strategy is similar to \ci{BM, DKRS, DKS}).
In particular, in \ci{DKRS} one establishes,
in the case of a WE,
a long-time asymptotics
\be\la{lt}
U(t)Y_0=\Theta U_0(t)Y_0+\rho(t)Y_0,\,\,\,t>0.
\ee
Here $U(t)$ is the dynamical group of
the WE with variable coefficients,
  $U_0(t)$ corresponds to
`free' equation, with constant coefficients, and
$\Theta$ is a `scattering operator'.
In this paper, instead of (\re{lt}), we use
a dual representation:
\be\la{dsti}
U^\pr(t)\Psi=U^\pr_0(t)W\Psi+r(t)\Psi, \,\,\,t\ge 0.
\ee
Here  $U^\pr(t)$ is a 'formal adjoint' to the dynamical group of
Eqn (\re{1.3}), while  $U_0^\pr(t)$ corresponds to the
`free' equation, with $A_k(x)\equiv 0$.
The remainder $r(t)$ is small in mean:
\be\la{rem}E|\langle Y_0,r(t)\Psi\rangle |^2\to 0,
\,\,\,t\to\infty.\ee
This version of scattering theory is essentially based on
Vainberg's bounds for the local energy decay (see \ci{V74,V89}).
\medskip\\
{\bf Remark} i)
In \ci{DKRS} we deduce asymptotics
(\re{lt}) from its primal counterpart (\re{dsti}).
In this paper we do not
analyse connections between
(\re{dsti}) and (\re{lt}).
\\
ii)
It is useful to
comment on the difference between two versions of scattering theory
produced for the WE and KGE.
In the first theory,
the remainders $\rho(t)$ and $r(t)$
are small a. s., while in the second theory,
developed in this paper,
 $r(t)$ is small in mean  (see  (\re{rem})).
Such a difference is related to a slow   (power)
decay of solutions to the KGE.
\medskip

The main result of the paper is stated in Section 2
(see Theorem A).
Sections 3 - 8 deal with the case of constant coefficients:
the main statement  is given in Section 3 (see Theorem B),
the relation to CGDs is discussed in Section 4,
 the compactness (Property {\bf I}) is established  in Section 5,
convergence (\re{corf}) in Section 6,
and convergence (\re{2.6i}) in Sections 7, 8.
In Section 9 we  check the Lindeberg condition
needed for convergence to a Gaussian limit.
In Section 10 we discuss the infinite energy
version of the scattering theory, and
in Section 11 convergence (\re{1.2}).
In Appendix A we collected  FT-type calculations.
Appendix B concerns with a formula on generalised GPMs on
Sobolev's spaces.
\bigskip\\
{\bf Acknowledgements}~~
The authors thank~
 V.I.~Arnold, A.~Bensoussan, I.A.~Ibragimov,
H.P.~McKean,  J.~Lebowitz, A.I.~Shnirelman,
H.~Spohn, B.R.~Vainberg and M.I.~Vishik for
fruitful  discussions and remarks.

\setcounter{equation}{0}
 \section{Main results}
\subsection{Notation}

We assume that  functions $A_k(x)$ in (\re{1.3})
satisfy the following conditions:
 \medskip\\
 {\bf E1.} $A_j(x)$ are real  $C^\infty$-functions.\\
 {\bf E2.} $A_j(x)=0$  for $|x|>R_0,$ where
 $R_0<\infty.$ \\
 {\bf E3.}
$\ds{\frac{\pa A_1}{\pa x_2}\not\equiv \frac{\pa A_2}{\pa x_1}}$
if $n=2.$
 \medskip\\

Assume that the initial date $Y_0$
belongs to the phase space ${\cal H}$ defined below.
\begin{definition}                 \la{d1.1}
$ {\cal H} \equiv H_{\rm loc}^1(\R^n)\oplus H_{\rm loc}^0(\R^n)$
is the Fr\'echet space
of pairs $Y(x)\equiv(u(x),v(x))$
of complex  functions $u(x)$, $v(x)$,
endowed with  local energy seminorms
\beqn                              \la{semin}
\Vert Y\Vert^2_{R}= \int\limits_{|x|<R}
\Big(
|v(x)|^2+
|\nabla u(x)|^2+m^2|u(x)|^2\Big) dx<\infty,
~~\forall R>0.
\eeqn
\end{definition}

 Proposition \re{p1.1} follows from
\ci[Thms V.3.1, V.3.2]{Mikh}) as the speed of
propagation for  Eqn (\re{1.3})
 is finite.
 \begin{pro}    \la{p1.1}
i)
For any  $Y_0 \in {\cal H}$
 there exists  a unique (generalised) solution
$Y(t)\in C(\R, {\cal H})$
 to (\re{CP}).
\\
ii) For any    $t\in \R$
the operator $U(t):Y_0\mapsto  Y(t)$
 is continuous in ${\cal H}$.
\end{pro}

Let us choose a function $\zeta(x)\in C_0^\infty(\R^n)$ with $\zeta(0)\ne 0$.
Denote by $H^s_{\rm loc}(\R^n),$ $s\in \R,$  the local Sobolev spaces,
i.e. the Fr\'echet spaces
of distributions $u\in D'(\R^n)$ with finite seminorms
\be\la{not}
\Vert u\Vert _{s,R}:= \Vert\La^s\Big(\zeta(x/R)u\Big)\Vert_{L^2(\R^n)},
\ee
where $\La^s v:=F^{-1}_{k\to x}(\langle k\rangle^s\hat v(k))$,
$\langle k\rangle:=\sqrt{|k|^2+1}$, and $\hat v:=F v$ is the FT
of a tempered distribution $v$. For $\psi\in D$ define
$F\psi ( k)= \ds\int e^{i k\cdot x} \psi(x) dx.$

\begin{definition}\la{d1.2}
For $s\in\R$  denote
$
{\cal H}^{s}\equiv H_{\rm loc}^{1+s }(\R^n)
\oplus H_{\rm loc}^{s }(\R^n).
$
\end{definition}

Using standard techniques of pseudodifferential operators
(see, e.g. \ci{H3}) and Sobolev's Theorem, it is possible to prove
 that
 ${\cal H}^0={\cal H}\subset {\cal H}^{-\ve }$ for every $\ve>0$,
and the embedding  is compact.

\subsection{Random solution. Convergence to equilibrium}

Let $(\Om,\Si,P)$ be a probability space
with expectation $E$
and ${\cal B}({\cal H})$ denote the Borel $\si$-algebra
in ${\cal H}$.
We assume that $Y_0=Y_0(\om,\cdot)$ in (\re{CP})
is a measurable
random function
with values in $({\cal H},\,{\cal B}({\cal H}))$.
In other words, $(\om,x)\mapsto Y_0(\om,x)$ is a measurable
 map
$\Om\times\R^n\to\C^2$ with respect to the
(completed)
$\si$-algebras
$\Si\times{\cal B}(\R^n)$ and ${\cal B}(\C^2)$.
Then, owing to Proposition \re{p1.1},
$Y(t)=U(t) Y_0$ is again a measurable  random function
with values in
$({\cal H},{\cal B}({\cal H}))$.
We denote by $\mu_0(dY_0)$ a probability measure
on ${\cal H}$
giving
the distribution of the  $Y_0$.
Without loss of generality,
 we assume $(\Om,\Si,P)=
({\cal H},{\cal B}({\cal H}),\mu_0)$
and $Y_0(\om,x)=\om(x)$ for
$\mu_0(d\om)\times dx$-almost all
$(\om,x)\in{\cal H}\times\R^n$.

\begin{definition}
$\mu_t$ is a  probability measure on ${\cal H}$
which gives
the distribution of $Y(t)$:
\begin{eqnarray}\la{1.6}
\mu_t(B) = \mu_0(U(-t)B),\,\,\,\,
 B\in {\cal B}({\cal H}),
\,\,\,   t\in \R.
\eeqn
\end{definition}

Our main goal is to derive
 the weak convergence of the measures $\mu_t$
in the
Fr\'echet space  ${\cal H}^{-\ve }$ for each $\ve>0$,
\be\la{1.8}
\mu_t\,\buildrel {\hspace{2mm}{\cal H}^{-\ve }}\over
{- \hspace{-2mm} \rightharpoondown }
\, \mu_\infty,\quad t\to \infty,
\ee
where $\mu_\infty$ is a limiting measure on the space
${\cal H}$. This means the convergence
 \be\la{1.8'}
 \int f(Y)\mu_t(dY)\rightarrow
 \int f(Y)\mu_\infty(dY),\quad t\to \infty,
 \ee
for any bounded continuous functional $f$
on ${\cal H}^{-\ve }$.
Recall that we identify $\C\equiv\R^2$ and
$\otimes$ stands for tensor product of real vectors.
Denote $M^2=\R^2\otimes\R^2$.

\begin{definition}
The CFs of the measure $\mu_t$ are defined by
\be\la{qd}
Q_t^{ij}(x,y)\equiv E \Big(Y^i(x,t)\otimes
{Y^j(y,t)}\Big),~~i,j= 0,1, ~~~~{\rm for~~almost~~all}~~
x,y\in\R^n\times\R^n,\ee
assuming that the expectations in the RHS are finite.
\end{definition}
We set ${\cal D}=D\oplus D$, and
$\langle Y,\Psi\rangle
=\langle Y^0,\Psi^0\rangle +\langle Y^1,\Psi^1\rangle$
for $Y=(Y^0,Y^1)\in {\cal H}$ and $
\Psi=(\Psi^0,\Psi^1)\in  {\cal D}$.
For a probability  measure $\mu$ on  ${\cal H}$,
denote by $\hat\mu$
the characteristic functional (the FT)
$$\hat \mu(\Psi )\equiv\int\exp(i\langle Y,\Psi \rangle )\,
\mu(dY),\,\,\,\Psi\in{\cal D}.$$
A probability measure $\mu$ is called a GPM (of mean zero) if
its characteristic functional has the form
$$\ds\hat { \mu} (\Psi ) =  \ds \exp\{-\fr12
{\cal Q}(\Psi , \Psi )\},\,\,\,\Psi \in {\cal D},$$
where ${\cal Q}$ is a real nonnegative quadratic form in
${\cal D}$. A measure $\mu$ is called translation-invariant if
$$\mu(T_h B)= \mu(B),\,\,\, B\in{\cal B}({\cal H}),
\,\,\,\, h\in\R^n,$$
where $T_h Y(x)= Y(x-h)$, $x\in\R^n$.

\subsection{Mixing condition}
Let $O(r)$ denote the set of all pairs of open bounded subsets
${\cal A},\>{\cal B}\subset \R^n$ at distance
dist$({\cal A},\,{\cal B})\geq r$ and $\sigma ({\cal A})$
the $\sigma $-algebra  in ${\cal H}$ generated by the
linear functionals $Y\mapsto\, \langle Y,\Psi\rangle $
where  $\Psi\in  {\cal D}$
with $ \supp \Psi \subset {\cal A}$.
Define the
Ibragimov-Linnik mixing coefficient
of a probability  measure  $\mu_0$ on ${\cal H}$
by (cf \ci[Dfn 17.2.2]{IL})
\be\la{ilc}
\varphi(r)\equiv
\sup_{({\cal A},{\cal B})\in O(r)} \sup_{
\ba{c} A\in\si({\cal A}),B\in\si({\cal B})\\ \mu_0(B)>0\ea}
\fr{| \mu_0(A\cap B) - \mu_0(A)\mu_0(B)|}{ \mu_0(B)}.
\ee
\begin{definition}\la{dmix}
 The measure $\mu_0$ satisfies the strong, uniform
Ibragimov-Linnik mixing condition if
\be\la{1.11}
\varphi(r)\to 0,\,\quad r\to\infty.
\ee
\end{definition}
Below, we  specify the rate of decay of $\varphi$
(see Condition {\bf S3}).

\subsection{Main assumptions and results}
We assume that measure $\mu_0$
has the following properties {\bf S0--S3}:
\bigskip\\
{\bf S0}
$\mu_0$ has zero expectation value,
\be
EY_0(x)  \equiv  0,\,\,\,\,\,\,x\in\R^n.
\la{1.10'}
\ee
{\bf S1} $\mu_0$ has translation-invariant CFs.
 i.e. Eqn
(\re{1.9'}) holds
for almost all $x,y\in\R^n$.
\\
{\bf S2}  $\mu_0$ has a finite mean energy density, i.e.
Eqn (\re{med}) holds.\\
{\bf S3}
 $\mu_0$ satisfies the strong uniform
Ibragimov-Linnik mixing condition with
 \be\la{1.12}
 \ov\varphi\equiv\ds\int\limits
_0^\infty r^{n-1}\varphi^{1/2}(r)dr <\infty.
 \ee
\medskip

Define, for almost all $x,y\in\R^n$,
the  matrix  $Q_{\infty}(x,y)\equiv
\Big(Q_{\infty}^{ij}(x,y)\Big)_{i,j=0,1}$
by
 \beqn\la{1.13}
 Q_{\infty}(x,y)\equiv
 \frac{1}{2} \left(
 \ba{lr}
 (q_0^{00}+{\cal P} * q_0^{11})(x-y) &
 (q_0^{01}-q_0^{10})(x-y) \\
 \hspace{8mm}
 (q_0^{10}-q_0^{01})(x-y) & (q_0^{11}-(\De-m^2) q_0^{00})(x-y)
 \ea     \right).
 \eeqn
Here   ${\cal P} (z)$ is
the fundamental solution for the
operator $-\De+m^2$, and
$*$ stands for  the convolution of generalized
functions.
We show below that $q_0^{11}\in L^2(\R^n)$
(see (\re{4.7})). Then
the convolution ${\cal P}*q_0^{11}$ in (\ref{1.13}) also belongs
to  $L^2(\R^n)$.

Let $ H =L^2(\R^n)\oplus H^1(\R^n)$  denote the space
of complex valued functions $\Psi=(\Psi_0,\Psi_1)$
  with a finite norm
\be\la{1.20}
\Vert\Psi\Vert_{ H}^2= \int\limits_{\R^n}
(|\Psi_0(x)|^2+|\nabla\Psi_1(x)|^2+|\Psi_1(x)|^2)\,dx
<\infty.
\ee
Denote by
${\cal Q}_{\infty}$ a real quadratic form
in $H$ defined by
\be\la{qpp}
{\cal Q}_\infty (\Psi, {\Psi})=\sum\limits_{i,j=0,1}~
\int\limits_{\R^n\times\R^n}
\Big(Q_{\infty}^{ij}(x,y)\Psi_i(x),\Psi_j(y)\Big)
dx~dy,
\ee
where $\Big(\cdot,\cdot\Big)$ stands for real scalar product in
$\C^2\equiv\R^4$.
 The form ${\cal Q}_\infty$ is continuous in $H$ 
by Corollary \ref{coro}.
\medskip\\
{\bf Theorem A}
{\it
    Let $n\ge 2$, $m>0$,
and assume that {\bf E1--E3}, {\bf S0--S3} hold.
 Then \\
i) The  convergence in (\re{1.8}) holds
for any $\ve>0$.\\
ii) The limiting measure
$\mu_\infty $ is a GPM on ${\cal H}$.\\
iii)~The  characteristic functional of $\mu_\infty $ has the form
$$
\ds\hat { \mu}_\infty (\Psi ) = \exp
\{-\fr{1}{2} {\cal Q}_\infty (W \Psi, {W \Psi})\},
\,\,\,
\Psi \in {\cal D},
$$
where  $W: {\cal D}\to  H$
is
 a linear continuous operator.
}


\subsection{Remarks on   conditions
on the  initial measure}
i)  The (rather strong) form of mixing in Definition \re{dmix}
is motivated by two facts:
(a) it greatly simplifies the forthcoming arguments,
(b) it allows us to produce an `optimal' (most slow)
decay of $\varphi$ indicating natural limits of Bernstein's
room-corridor method.
Condition  (\re{ilc}) can be easily verified for GPMs
with finite-range  dependence and their images under `local' maps
${\cal H}\to {\cal H}$. See the examples
in Section 2.6 below.\\
ii) The {\it uniform} Rosenblatt mixing condition
\ci{Ros} also suffices, together with a higher
power $>2$  in the bound (\re{med}): there exists $\de >0$ such that
$$
~~~~~~~~~~~~
~~~~~~~~~~~
E \Big(
\vert v_0(x)\vert^{2+\de}+
\vert \nabla u_0(x)\vert^{2+\de}+m^2\vert u_0(x)\vert^{2+\de}
\Big)
<\infty.~~~~~~~~~~~~~~~~~~~~~~~~~~~~~~~~\eqno{(1.4')}
$$
Then  (\re{1.12}) requires a modification:
$$
~~~~~~~~~~~~~~~
~~~~~~\int _0^\infty\ds r^{n-1}\al^{p}(r)dr <\infty,\,\,\,\,
\mbox{where}\,\,\,\,
p=\min(\fr\de{2+\de}, \fr 12),~~
~~~~~~~~~~~~~~~~~~~~~~~~~~~~~~~~~~~~~~\eqno{(2.10')}
$$
where $\al(r)$ is the  Rosenblatt
mixing coefficient  defined
as in  (\re{ilc}) but without $\mu(B)$ in the denominator.
The statements of Theorem A and their
proofs remain essentially unchanged, only Lemma \re{il} requires
a suitable modification \ci{IL}.

\subsection{Examples of initial measures with mixing condition}
\subsubsection{Gaussian measures}

In this section we construct initial GPMs $\mu_0$
satisfying {\bf S0 -- S3}.
Let $\mu_0$ be a GPM in the space ${\cal H}$
with the characteristic functional
\be\la{2.5}
\hat \mu_0(\Psi )  \equiv  E\exp(i\langle Y,\Psi \rangle )
= \exp\{-\fr{1}{2}{\cal Q}_0(\Psi ,\Psi )\},~~
\Psi  \in  {\cal D}.
\ee
Here ${\cal Q}_0$ is a real nonnegative quadtratic form with an
integral kernel
$(Q_0^{ij}(x,y))_{i,j=0,1}$.
Let
\be\la{2.5'}
Q_0^{ij}(x,y)\equiv q_0^{ij}(x - y),
\ee
for any $i,\,j$, where the function
$q_0^{ij}\in C^2 (\R^n)\otimes M^2$ has a compact support. Then
{\bf S0}, {\bf S1} and {\bf S2}
are satisfied;
{\bf S3}  holds
with $\varphi(r)\equiv 0$ for $r\geq r_0$
if $q_0^{ij}(z)\equiv 0$ for $| z| \geq r_0$.
For a given matrix function
$\Big(q_0^{ij}(z)\Big)$
such a measure exists in the space ${\cal H}$
iff the corresponding FT
is a nonnegative matrix-valued measure:
$\Big(\hat q_0^{ij}( k)\Big) \geq 0$, $ k\in\R^n$, \ci[Thm V.5.1]{GS}.
For example, all these conditions hold
if $\hat q_0^{ij}( k)=D_i\de^{ij}f( k_1)\cdot\dots\cdot f( k_n)$
with $D_i\ge 0$ and
$$
f(z)=
\left(\fr{1-\cos(r_0 z/\sqrt{n})}{z^2}\right)^2, \,\,\,z\in\R.
$$
\subsubsection{Non-Gaussian measures}
Now choose  a pair of odd  functions
 $f^0,\,f^1\in C^1 (\R),$ with
bounded first derivatives.
Define $\mu_0^*$ as the distribution of the random function
$
(f^0(Y^0(x)),  f^1(Y^1(x))),
$
where $(Y^0,Y^1)$ is a random function
with a Gaussian distribution $\mu_0$ from the previous
example.
Then {\bf S0-S3} hold for $\mu_0^*$
with a
mixing coefficient
$\varphi^*(r)\equiv 0$
 for $r\geq r_0$.
Measure $\mu_0^*$
is not Gaussian
if $D_i>0$ and
 the functions $f^i$ are bounded and nonconstant.

\setcounter{equation}{0}
 \section{ Equations with  constant coefficients}

In Sections 3-9 we assume that coefficients
 $A_k(x)\equiv 0$. Problem (\ref{1.3}) then becomes
 \beqn\la{2.1}
\left\{
\ba{l}
 \ddot u(x,t) = \Delta u(x,t) - m^2u(x,t),~~
~ t\in \R,\\
 u|_{t=0} = u_0(x),~~ \dot u|_{t=0} = v_0(x).
\ea
\right.
\eeqn
As in (\ref{CP}), we  rewrite (\ref{2.1}) in the form
\be\la{2'}
\dot Y(t)={\cal A}_0Y(t),\,\,\,t\in\R;\,\,\,\,Y(0)=Y_0.
\ee
Here  we denote
\be\la{A0}
{\cal A}_0=\left(
 \begin{array}{cc}
0 & 1\\
A_0 & 0
\end{array}\right),
\ee
where
 $A_0=\De-m^2$.
Denote by $U_0(t),$ $t\in\R,$ the dynamical group for problem
(\ref{2'}), then $Y(t)=U_0(t)Y_0$.
 The following proposition is well-known
and is proved by a standard partial integration.
 \begin{pro}\la{p2.1}
Let $Y_0=(u_0,v_0)\in {\cal H}$,
and $Y(\cdot,t)=(u(\cdot,t),\dot u(\cdot,t))
\in C(\R, {\cal H})$ is the
 solution to (\ref{2.1}).
Then
the following energy bound holds: for
 $R>0$ and $t\in\R$,
 \beqn\la{2.4}
\!\!
\int\limits_{|x|<R}\!\!\Big(|\dot u(x,t)|^2\!+\!|\nabla u(x,t)|^2\!+
\!m^2|u(x,t)|^2
\Big)dx
  \le \! \!
\int\limits_{|x|<R+|t|}\!\!\!\Big(|v_0(x)^2|\!+\!|\nabla
 u_0(x)|^2\!+\!m^2|u_0(x)|^2\Big)dx.
 \eeqn
 \end{pro}

Set  $\mu_t(B)=\mu_0(U_0(-t)B)$,
$B\in {\cal B}({\cal H})$, $t\in\R$. Then
 our main result for  problem (\re{2'}) is
\medskip\\
{\bf Theorem B}
{\it
    Let $n\ge 1$, $m>0$, and Conditions
 {\bf S0--S3} hold.
 Then the conclusions  of Theorem A
  hold with $W=I$,
and limiting measure $\mu_\infty$ is translation-invariant.}
\medskip\\
Theorem B can be deduced
 from Propositions  \re{l2.1} and \re{l2.2} below,
 by the same arguments as in
\ci[Thm XII.5.2]{VF}.

\begin{pro}\la{l2.1}
  The family of measures $\{\mu_t, t\in \R\}$,
 is weakly compact in   ${\cal H}^{-\ve }$ with any $\ve>0$,
and the bounds hold:
\be\la{p3.1}
\sup\limits_{t\ge 0}
 E \Vert U_0(t)Y_0\Vert^2_R <\infty,\,\,\,\,R>0.
\ee
 \end{pro}
 \begin{pro}\la{l2.2}
  For every $\Psi\in {\cal D}$,
 \be\la{2.6}
 \hat \mu_t(\Psi )\equiv\int \exp(i\langle Y,\Psi\rangle )\mu_t(dY)
 \rightarrow \exp\{-\fr 12{\cal Q}_\infty( \Psi,\Psi)\}.
 \,\,\,t\to\infty.
 \ee
 \end{pro}
Propositions \re{l2.1} and  \re{l2.2}  are proved in
Sections 5 and 7-9, respectively.
We will use repeatedly the FT (\re{hatA}) and  (\re{tidtx})
from Appendix A.


\section{Relation to CGDs}
\setcounter{equation}{0}
In this section we discuss how our results are related to CGDs.
We restrict consideration to the case of Eqn
(\re{1.3}) with constant coefficients
and to the translation-invariant isotropic case.
The CGD $g_T$ with the absolute temperature $T\ge 0$
is defined formally by
\be\la{5}
g_T(du\times dv)= \frac 1Z
\ds~ e^{-\ds \fr H  T}
\prod_{x}
du(x)dv(x),
\ee
where $\ds H:=\fr 12\int\Big(|v(x)|^2+|\nabla u(x)|^2+m^2|u(x)|^2\Big)
 dx$,
and $Z$ is a normalisation constant.

To make the definition rigorous,
let us  introduce a  scale of weighted  Sobolev's spaces
$H^{s,\al}(\R^n)$
with  arbitrary  $s,\al \in\R$.
We use notation  (\re{not}).
\begin{definition}  i)
$H^{s,\al}(\R^n)$
is the complex Hilbert space of the distributions
$w\in S'(\R^n)$ with the finite norm
\be\la{ws}
\Vert w\Vert_{s,\al}\equiv
\Vert \langle x\rangle^{\al}\Lambda^s
w \Vert_{L_2(\R^n)}<\infty.
\ee
ii)
 ${\cal H}^{s,\al }$ is the Hilbert space
of the pairs $Y=(u,v)\in H^{1+s,\al }(\R^n)
 \oplus H^{s,\al}(\R^n)$
with the norm
\be\la{wsb}
\brr Y\brr_{s,\al}\equiv\Vert u\Vert_{1+s,\al}+\Vert v\Vert_{s,\al}.
\ee
\end{definition}

Note that  ${\cal H}^{\ov s,\ov \al }\subset {\cal H}^{s,\al }$ if
$\ov s<s$ and $\ov \al<\al$,
and this embedding  is compact.
These facts follow by standard methods of pseudodifferential
operators and
Sobolev's Theorem (see, e.g. \ci{H3}).

Now we can define the CGDs  rigorously:
$g_T$ is a GPM
on a space ${\cal H}^{s,\al}$, $s,\al<-n/2$,
with the CFs
\be\la{4}
g_T^{00}(x-y)= T{\cal P}(x-y),~~
g_T^{11}(x-y)= T\delta(x-y),~~
g_T^{01}(x-y)= g_T^{10}(x-y)= 0.
\ee
By Minlos Theorem  \ci[Thm V.5.1]{GS},
such a measure exists on ${\cal H}^{s,\al}$
with $s,\al<-n/2$ as, {\it formally} (see Appendix B),
\be\la{Min}
\int\brr Y\brr_{s,\al}^2~g_T(dY)<\infty.
\ee
Measure $g_T$ is stationary for the KGE,
as its CFs are stationary; the last fact
follows from formulas (\re{tidt}), (\re{hatA}).
Also,
$g_T$ is translation invariant, so {\bf S1} holds.
Condition {\bf S2} fails since
the `mean energy density'
$g_T^{11}(0)-\De g_T^{00}(0)+m^2g_T^{00}(0)$
is infinite; this gives an `ultraviolet divergence'.
Mixing condition  {\bf S3} holds
due to
an exponential decay of the
${\cal P}(z)$.
The convergence of type
(\re{1.1}) holds for
 initial measures $\mu_0$ that are absolutely continuous
with respect to the CGD $g_T$, and
the limit measure coincides with $g_T$.
This mixing property  (and even $K$-property)
can be proved by using well-known methods developed
for Gaussian
processes \ci{CFS}, and we do not discuss it here.
\medskip\\
{\bf Remark}
Assumption {\bf S2}
implies that
$\mu_0({\cal H})=1$ and
hence
 $\mu_\infty({\cal H})=1$.
This excludes the case of a limiting
CGD as it is a generalised  GPM
not supported by ${\cal H}$.
However, it is possible
to extend  our results to a class
of generalised initial measures converging to CGDs.
For the case of constant coefficients
such an extension could be done by
smearing the
initial  generalised  field as the dynamics commutes
with  the averaging (cf. \ci{DKS}). For variable coefficients
such extension
requires a further work.
\medskip

To demonstrate the special role  of the CGDs we consider a family
of initial GPMs  $\mu_{0,r}$, $r\in (0,1]$,
satisfying {\bf S0}-{\bf S3},
with the radius of correlation $r$.
More precisely, suppose that the
corresponding CFs $q^{ij}_{0,r}$ have
the following properties {\bf G0}-{\bf G3}:\\
$$
\ba{lrl}
{\bf G0}&~   &~~~~~~~~~~~~~~~~~~~~~~~
q^{01}_{0,r}(z)=q^{01}_{0,r}(-z),\,\,\, z\in\R^n.~~~~~~~~~~~
~~~~~~~~~~~~\\
~&\\
{\bf G1}&~   &~~~~~~~~~~~~~~~~~~~~~~~
q^{11}_{0,r}(z)-\De q^{00}_{0,r}(z)+m^2 q^{00}_{0,r}(z)
=0,\,\,\,|z|\ge r.~~~~~~~~~~~~~~~~~~~~~~~~~\\
~&\\
{\bf G2}&\mbox{For some} &\!\!T>0,~~~~~~
\ds\fr 12
\ds~\int \Big(~q^{11}_{0,r}(z)-\De q^{00}_{0,r}(z)+m^2 q^{00}_{0,r}(z)~\Big)dz
\to T,\,\,\,r\to 0.~~~~~~~~~~~~~~~~~~~~~~~\\
~&\\
{\bf G3}&  & ~~~~~~~~~\sup\limits_{r\in (0,1]}
\ds \int \Big(|q^{11}_{0,r}(z)|\!+\!|\De q^{00}_{0,r}(z)|\!+\!m^2
|q^{00}_{0,r}(z)|
\Big)dz
<\!\!\infty.~~~~~~~~~~~~~~~~~~~~~
\ea$$
Note that {\bf G0} means a symmetry relation
$E u_0(x) v_0(y)=E u_0(y) v_0(x)$
that holds for an isotropic measure where the CFs
depend only on $|x-y|$.
Examples of such family will be provided later.

Properties {\bf G0}-{\bf G3} imply conditions
{\bf S0}-{\bf S3} for the initial measures $\mu_{0,r}$.
Therefore, Theorem B implies the convergence
$\mu_{t,r}\,\buildrel {\hspace{2mm}{\cal H}^{-\ve }}\over
{- \hspace{-2mm} \rightharpoondown }
\, \mu_{\infty,r},\quad t\to \infty$,
of type (\re{1.8}). The following proposition means that
the limiting measure $\mu_{\infty,r}$ is close to CGD
$g_T$ on the Sobolev space of distributions
${\cal H}^{s,\al}$ with  $s,\al<-n/2$.

\begin{pro}\la{pG}
Let Conditions  {\bf G0}-{\bf G3} hold.
Then corresponding limiting
measures $\mu_{\infty,r}$
are concentrated on any space
${\cal H}^{s,\al}$ with  $s,\al<-n/2$ and
weakly converge to CGD $g_T$ on the space
${\cal H}^{s,\al}$:
\be\la{1.8gT}\mu_{\infty,r}\,\buildrel
 {\hspace{2mm}{\cal H}^{s,\al }}\over
{- \hspace{-2mm} \rightharpoondown}\,
g_T,\,\,\,\, r\to 0.\ee
\end{pro}
{\bf Proof} The convergence follows by the same arguments as in
\ci{VF} from two facts (cf. Propositions \re{l2.1}, \re{l2.2}):
for any $\ov s,\ov\al$ with  $s<\ov s<-n/2$ and  $\al<\ov\al<-n/2$,
$$\ba{ll}
{\bf (I)}~~~~~~~~~~~~~~    & \sup\limits_{r\in (0,1]}
\ds\int
\brr Y\brr_{\ov s,\ov\al}^2\mu_{\infty,r}(dY)<\infty.
~~~~~~~~~~~~~~~ ~~~~~~~~~~~~~~~~~~~~~~~~~    \\
~\\
{\bf (II)}~~~~~~~~~{\rm For~~~}\Psi\in{\cal D},~~~~~
 & {\cal Q}_{\infty,r}(\Psi,\Psi)\to
{\cal G}_T(\Psi,\Psi),\,\,r\to 0,
~~~~~~~~~~~~~~~ ~~~~~~~~~~~~~~~ ~~~~~~~~~
\ea$$
where ${\cal Q}_{\infty,r}$ is the
quadratic form with the  integral kernel $\Big(q^{ij}_\infty(x-y)\Big)$,
and ${\cal G}_T$ corresponds to $\Big(g^{ij}_T(x-y)\Big)$.
It is important that the embedding
${\cal H}^{\ov s,\ov \al}\subset{\cal H}^{s,\al}$  is compact.
Property {\bf (I)}  can be checked with the help of the formula
(\re{wsmu}) and by
 using the Parseval identity:
$$
\ba{rcl}
\ds\int
\brr Y\brr_{\ov s,\ov\al}^2\mu_{\infty,r}(dY)&\!\!\!=&\!\!\!
\ds \fr{C(\ov\al)}{(2\pi)^n}
\int\Big(\langle k\rangle^{2\ov s}~ \tr\hat q^{11}_{\infty,r}(k)+
\langle k\rangle^{2(1+\ov s)}~ \tr\hat q^{00}_{\infty,r}(k)\Big)dk
\nonumber\\
~\nonumber\\
&\!\!\!=&\!\!\!C(\ov\al)\ds\int
\Big(f_1(z) ~ \tr q^{11}_{\infty,r}(z)+
f_2(z)  (-\De+m^2)\tr q^{00}_{\infty,r}(z)\Big)dz,
\ea
$$
where $f_1(z)=\ds\fr 1{(2\pi)^n}\ds\int e^{-ikz}
\langle k\rangle^{2\ov s} dk$
and $f_2(z)=\ds\fr 1{(2\pi)^n}\ds\int e^{-ikz}
\fr{\langle k\rangle^{2(1+\ov s)}}{k^2+m^2} dk$.
More precisely, Property {\bf (I)}  follows  from
{\bf G3}  as  both  functions $f_j(x)$
are bounded and continuous for $\ov s<-n/2$.
Furthermore, {\bf G0} and (\re{1.13}) imply that
$q^{01}_{\infty,r}=q^{10}_{\infty,r}=0$, hence
\be\la{qinf}
{\cal Q}_{\infty,r}(\Psi,\Psi)=\int\Big( q^{00}_{\infty,r}(x-y)\Psi^0(x),
{\Psi^0(y)}\Big)dxdy
+\int\Big( q^{11}_{\infty,r}(x-y)\Psi^1(x),{\Psi^1(y)}\Big)dxdy.
\ee
{\bf G1} and {\bf G2}  together imply that
$$
q^{11}_{\infty,r}(x-y)\to T\de(x-y),\,\,\,r\to 0.
$$
Then (\re{1.13}) implies
$$
q^{00}_{\infty,r}={\cal P}*q^{11}_{\infty,r}
\to T {\cal P},\,\,\,\,r\to 0.
$$
Therefore, Property {\bf (II)} follows from (\re{qinf}):
the justification follows easily in the FT space.
The convergence of the covariance {\bf (II)}
provides the convergence of the measures (\re{1.8gT})
as all the measures are Gaussian.
\hfill$\Box$
\medskip\\
{\bf Example}
Consider an initial mesure $\mu_0$
constructed in Example in Section 2.6.1.
It
satisfies  Assumptions  {\bf S0}-{\bf S3}
 and {\bf G0}. Furthermore,
$
 q_0^{00}(z)=q_0^{11}(z)=0,\,\,\,|z|\ge 1
$ if we choose $r_0=1$.
Denote by $Y_0(x)=(Y_0^0(x),Y_0^1(x))$ a random function
with distribution $\mu_0$. Denote by $\mu_{0, r }$, $ r >0$,
the distribution of the random function
$\ds Y_{0, r }(x)=\ds( r ^{1-\nu}Y_0^0( r ^{-1}x),
 r ^{-\nu}Y_0^1( r ^{-1}x))$ where $\ds\nu= n/2$.
The corresponding CFs are
$
q_{0, r }^{ij}(z)= r ^{2-n-i-j}
q_0^{ij}( r ^{-1} z).
$
Then all Conditions {\bf G0}-{\bf G3} hold
with $T:=\ds\fr 14 \int\tr q_0^{11}(z)dz$.

\setcounter{equation}{0}
\section{Compactness of the family of measures $\mu_t$}
This section gives the proof of bound
(\re{p3.1}).
Proposition \re{l2.1} will follow then with the help of
the Prokhorov Theorem
 \cite[Lemma II.3.1]{VF} as in the proof of
 \ci[Thm XII.5.2]{VF}.
It is important that
 the embedding ${\cal H}\subset {\cal H}^{-\ve }$
 is compact,
by virtue of Sobolev's  Theorem, if $\ve>0$.
Set:
\be e_t\equiv
E\Big(| \dot u(x,t)| ^2 +|\nabla u(x,t)| ^2+m^2| u(x,t)|^2\Big),
\,\,\,\, x\in \R^n.                \la{3.2} \ee
The CFs of
measure $\mu_t$ are translation invariant due to
condition {\bf S1}. Hence, taking expectation in  (\re{2.4}),
we get by {\bf S2},
\be\la{3.4}
e_t |B_R|\le  e_0  |B_{R+t}|<\infty.\ee
Here $B_R$ is the ball $| x| \le R$ in $\R^n$, and
$| B_R| $ is its volume.
Taking $R\rightarrow \infty $ we derive from (\re{3.4}) that
$e_t\le e_0$: in fact, the reversibility implies then
$e_t= e_0$ (the mean energy density conservation).
Hence, taking expectation in
(\re{1.5}), we get (\re{p3.1}):
$$
~~~~~~~~~~~~~~~~~~~~~~~~~~~~~~~~~~~~~~
~~~~~~~~~~~~E\Vert U_0(t) Y_0\Vert_R^2=
e_0 |B_R|<\infty.~~~~~~~~~~~~~~~~~~~~~~~~~~
~~~~~~~~~~~~~~~~\Box
$$
\begin{cor}\la{Q}
Bound (\re{p3.1})
implies the convergence of the integrals in (\re{qd}).
\end{cor}
\setcounter{equation}{0}
\section{Convergence of the covariance functions}
In this section we check the convergence
of the CFs of measures $\mu_t$
with the help of the FT.
This convergence is used in Section 8.

\subsection{Mixing in terms of the spectral density}
The next proposition gives the mixing property in terms of the
FT $\hat q^{ij}_0$ of the initial CFs $q^{ij}_0$.
Assumption {\bf S2} implies that  $q^{ij}_0(z)$
is a measurable bounded  function.
Therefore, it belongs to the Schwartz space
of tempered distributions as well as its FT.
\begin{pro} \la{l4.1}
Let the assumptions of Theorem B hold.
Then $\hat q^{ij}_0\in  L^1(\R^n)\otimes M^2$,~ $\forall i,j$.
\end{pro}
{\bf Proof} {\it Step 1}~ First, let us prove that
\beqn
\pa^{\gamma}q^{ij}_0(z)&\in&
L^p(\R^n)\otimes M^2,~~\,\,\,\,\,\,\,\mbox{~}
 p\ge 1,\,\,\,|\gamma|\le 2-i-j.\la{4.7}
\eeqn
Conditions {\bf S0}, {\bf S2} and {\bf S3} imply. by
\ci[Lemma 17.2.3]{IL} (see Lemma \re{il} i) below), that
\be\la{4.9}
|\pa^{\gamma}q^{ij}_0(z)|\le Ce_0\varphi^{1/2}(|z|),~~ z\in\R^n.
\ee
Mixing coefficient $\varphi$ is bounded, hence
 (\ref{4.9}) and (\ref{1.12}) imply (\ref{4.7}):
\be\la{qp}
\int
|\pa^{\gamma}q^{ij}_0(z)|^p\,dz\le Ce_0^p\int\limits_{\R^n}
\varphi^{p/2}(|z|)\,dz\le  C_1e_0^p
 \int _0^\infty r^{n-1}\varphi^{1/2}(r)dr <\infty.
\ee
{\it Step 2}~ By Bohner's theorem,
$\hat q_0\equiv(\hat q^{ij}_0(k))dk$ is a
complex positive-definite
matrix-valued measure
on $\R^n$,
 and {\bf S2} implies that the total measure
 $\hat q_0(\R^n)$ is finite.
On the other hand, (\ref{4.7}) with $p=2$  implies that
$\hat q^{ij}_0\in L^2(\R^n)\otimes M^2$.
\hfill$\Box$
\subsection{Proof of convergence of covariance functions}

Formulas (\ref{tidtx}),  (\ref{hatA}) and
Proposition \re{l4.1} imply  for example,
\beqn
&& q_t^{00}(x-y):=
E\Big(u(x,t)\otimes{u(y,t)}~\Big)
\la{4.6}\\
~\nonumber\\
&&=\fr 1{(2\pi)^n}
\int
e^{-i k(x-y) }
\Bigl[ \frac{1+\cos 2\om t}{2}
\hat q_0^{00}( k)+\frac{\sin 2 \om t}{2\om}
(\hat q_0^{01}( k)+\hat q_0^{10}( k))
+
\frac{1-\cos 2\om t}{2\om^2}
\hat q_0^{11}( k)\Bigr]\,d k,\nonumber
\eeqn
where the integral converges and
   defines a continuous  function
determined for {\it all}  $x,y\in\R^n$. Similar integrals
give a convenient
modification for all functions $q^{ij}_t(x-y)$, which we will work with.

\begin{pro}\la{p4.1}
Covariance functions $q_t^{ij}(z)$, $i,j=0,1$,
converge for all $z\in\R^n$:
\be\la{4.4}
q_t^{ij}(z)\to q_{\infty}^{ij}(z),~~\mbox{ }~t\to\infty,
\ee
where functions $q_{\infty}^{ij}(z)$ are defined in
(\ref{1.13}).
\end{pro}
{\bf Proof.}
(\ref{4.6})  and   Proposition  \re{l4.1}
imply,
\be\la{4.5'}
q_t^{00}(z)
\to\frac{1}{2}\Bigl(q_0^{00}(z)+{\cal P} * q_0^{11}(z)\Bigr)
= q_\infty^{00}(z),~~
t\to\infty,
\ee
as the oscillatory integrals tend to zero by the Lebesgue--Riemann Lemma.
For other $i,j$ the proof is similar.\hfill$\Box$
\medskip

Note that ${\cal P} (z)\in L^1(\R^n)$. Therefore,
(\ref{4.7})  with $p=1$  and explicit formulas
(\ref{1.13})
imply the following
\begin{cor}\la{coro}
Functions $q^{ij}_\infty$ belong to
$L^1(\R^n)\otimes M^2$,~ $i,j=0,1$.
\end{cor}

\begin{remark}
{\rm A similar argument in the FT representation
 implies  compactness in Pro\-po\-si\-tion \ref{l2.1}.
We provided  an independent proof of the compactness in
Section  5 to show the relation
with energy conservation.
}
\end{remark}
\setcounter{equation}{0}
\section{Bernstein's argument for the Klein--Gordon equation}
In this and the subsequent section we develop
a version of
Bernstein's `room-corridor'
method.
We use the standard integral
representation  for solutions,
 divide the domain of integration
into `rooms' and `corridors' and evaluate
their contribution. As a result, the value $\langle U_0(t)Y_0,\Psi\rangle $
for $\Psi\in{\cal D}$
is represented
as the sum of
weakly dependent random variables.
We  evaluate  the variances  of these  random variables
which will be important in  next section.

First, we evaluate $\langle Y(t),\Psi\rangle $ in (\ref{2.6})
by using  duality arguments.
For  $t\in\R$,
introduce a `formal adjoint' operators
$U'_0(t)$, $U'(t)$
from  space ${\cal D}$ to a suitable space of distributions.
For example,
\be\la{def}
\langle Y,U'_0(t)\Psi\rangle =
\langle U_0(t)Y,\Psi\rangle ,\,\,\,
\Psi\in {\cal D},
\,\,\, Y\in {\cal H}.
\ee
 Denote $\Phi(\cdot,t)=U'_0(t)\Psi$. Then
(\ref{def}) can be rewriten as
\be\la{defY}
\langle Y(t),\Psi\rangle =\langle Y_0,\Phi(\cdot,t)\rangle,
\,\,\,\,t\in\R.
\ee
The adjoint groups
admit a convenient description.
Lemma \re{ldu} below displays that
the action of groups $U'_0(t)$, $U'(t)$ coincides,
respectively, with the action
of $U_0(t)$, $U(t)$, up to the order of the components.
In particular, $U'_0(t)$, $U(t)$ are continuous
groups of operators  ${\cal D}\to {\cal D}$.
\begin{lemma}\la{ldu}
For $\Psi=(\Psi^0,\Psi^1)\in {\cal D}$
\be\la{UP}
U'_0(t)\Psi= (\dot\phi(\cdot,t),\phi(\cdot,t)),\,\,\,\,\,
U'(t)\Psi= (\dot\psi(\cdot,t),\psi(\cdot,t)),
\ee
where $\phi(x,t)$ is the solution of Eq(\ref{2.1})
 with the initial date
$(u_0,v_0)=(\Psi^1,\Psi^0)$
and $\psi(x,t)$ is the solution of Eqn (\ref{1.3})
with the initial date $(u_0,v_0)=(\Psi^1,\Psi^0)$.
\end{lemma}
{\bf Proof}
Differentiating (\ref{def}) in $t$ with
$Y,\Psi\in {\cal D}$, we obtain
\be\la{UY}
\langle Y,\dot U'_0(t)\Psi\rangle =\langle
\dot U_0(t)Y,\Psi\rangle .
\ee
Group $U_0(t)$ has the generator (\ref{A0}).
The generator of $U'_0(t)$   is the conjugate operator
\be\la{A0'}
{\cal A}'_0=
\left( \begin{array}{cc} 0 & A_0 \\
1 & 0 \end{array}\right).
\ee
Hence, Eqn (\ref{UP}) holds with
 $\ddot \psi=A_0\psi$.
For the group $U'(t)$ the proof is similar.
 \hfill$\Box$\\
\medskip

Next we introduce a `room-corridor'  partition of  $\R^n$.
Given $t>0$,
choose
$d\equiv d_t\ge 1$ and
$\rho\equiv\rho_t>0$.
Asymptotical relations between $t$, $d_t$ and  $\rho_t$
are specified below.
Set $h=d+\rho$ and
\be\la{rom}
a^j=jh,\,\,\,b^j=a^j+d,
\,\,\,j\in\Z.
\ee
We call the slabs $R_t^j=\{x\in\R^n:~a^j\le x^n\le b^j\}$  `rooms'
and $C_t^j=\{x\in\R^n:~b^j\le x^n\le  a_{j+1}\}$  `corridors'.
Here  $x=(x^1,\dots,x^n)$,
$d$ is the width of a room, and
$\rho$  of a corridor.

Denote  by
 $\chi_r$ the indicator of the interval $[0,~ d]$ and
$\chi_c$ that of $[d,~ h]$ so that
$\sum_{j\in \Z}(\chi_r(s-jh)+\chi_c(s-jh))=1$ for
(almost all) $s\in\R$.
The  following  decomposition holds:
\be\la{res}
\langle Y_0,\Phi(\cdot,t)\rangle = \sum_{j\in \Z}
(\langle Y_0,\chi_r^j\Phi(\cdot,t)\rangle +\langle
Y_0,\chi_c^j\Phi(\cdot,t)\rangle ),
\ee
where $\chi_r^j:=\chi_r(x^n-jh)$ and
$\chi_c^j:=\chi_c(x^n-jh)$.
Consider
random variables
 $ r_t^j$, $ c_t^j$, where
\be\la{100}
r_t^j= \langle Y_0,\chi_r^j\Phi(\cdot,t)\rangle  ,~~
c_t^j= \langle Y_0,\chi_c^j\Phi(\cdot,t)\rangle ,~~~~~~j\in\Z.
\ee
Then  (\ref{res}) and (\ref{defY}) imply
\be\la{razli}
\langle U_0(t)Y_0,\Psi\rangle =\sum\limits_{j\in\Z}
(r_t^j+c_t^j).
\ee
The series  in (\ref{razli})
is indeed a finite sum.
 In fact,  (\ref{A0'}) and (\ref{Frep})   imply that
in the FT representation,
$\dot{\hat \Phi}(k,t)=\hat{\cal A}'_0 (k)\hat \Phi(k,t)$ and
$
\hat \Phi(k,t)=\hat{\cal G}'_t( k)
\hat\Psi(k).
$
Therefore,
\be\la{frep'}
\Phi(x,t)=\fr 1{(2\pi)^n}
\int\limits_{\R^n} e^{-ikx} \hat{\cal G}'_t( k)\hat\Psi( k)~d k.
\ee
This can be rewritten as a convolution
\be\la{conr}
\Phi(\cdot,t)={\cal R}_t*\Psi,
\ee
where ${\cal R}_t=F^{-1}\hat{\cal G}'_t$.
The support  supp$\hspace{0.5mm}\Psi\subset B_{\ov r}$ with an ${\ov r}>0$.
Then the  convolution representation (\ref{conr}) implies that
the support of the function $\Phi$ at $t> 0$
is a subset of an  `inflated future cone'
\be\la{conp}
{\rm supp}\hspace{0.5mm}\Phi\subset\{(x,t)\in\R^n\times{\R_+}:
~ |x|\le t+{\ov r}\}.
\ee
as ${\cal R}_t(x)$ is supported by the `future cone' $|x|\le t$.
The last fact follows from
general  formulas (see \ci[(II.4.5.12)]{EKS}), or from the
Paley-Wiener
Theorem (see, e.g. \ci[Thm II.2.5.1]{EKS}),
as $\hat{\cal R}_t(k)$ is an entire  function
of $k\in\C^n$  satisfying suitable bounds.
Finally, (\ref{100})  implies that
\be\la{1000}
r_t^j= c_t^j= 0
\,\,\,\quad{\rm     for     }\quad\quad \,jh+t< -{\ov r}\,\,\,
\quad\mbox{    or    }\quad\,\,jh-t>{\ov r}.
\ee
Therefore,  series (\re{razli}) becomes
a  sum
\be\la{razl}
\langle U_0(t)Y_0,\Psi\rangle =\sum\limits_{-N_t}^{N_t}
(r_t^j+c_t^j),\,\,\,\,\ds N_t\sim \fr th
\ee
as $h\ge 1$.
\begin{lemma}  \la{l5.1}
    Let $n\ge 1$, $m>0$, and {\bf S0--S3} hold.
The following bounds hold for $t>1$:
\be\la{106}
E|r^j_t|^2\le  C(\Psi)~d_t/t,
\,\,\,\,
E|c^j_t|^2\le C(\Psi)~\rho_t/t,\,\,\,\,\,\,\,\,j\in\Z.
\ee
\end{lemma}
{\bf Proof}
We discuss the first bound in (\ref{106}) only, the second
is done in a similar way.

{\it Step 1}
 Rewrite the left hand side  as the integral of
CFs. Definition (\ref{100}) and
Corollary \re{Q}  imply by Fubini's Theorem
that
 \be\la{100rq}
E|r_t^j|^2= \langle \chi_r^j(x^n)\chi_r^j(y^n)q_0(x-y),
\Phi(x,t)\otimes {\Phi(y,t)}~ \rangle.
\ee
The following bound holds true
(cf. \ci[Thm XI.17 (b)]{RS3}):
\be\la{bphi}
\sup_{x\in\R^n}|\Phi(x,t)| ={\cal O}(t^{-n/2}),
\,\,\,\,t\to\infty.
\ee
In fact, (\ref{frep'}) and  (\ref{hatA}) imply that
$\Phi$ can be written as the sum
\be\la{frepe}
\Phi(x,t)=\fr 1{(2\pi)^n}\sum\limits_{\pm}~~
\int\limits_{\R^n} e^{-i(kx\mp\om t)}
a^\pm(\om)
\hat\Psi( k)~d k,
\ee
where $a^\pm(\om)$
is a matrix whose entries
are linear functions of $\om$ or $1/\om$.
Let us prove the asymptotics (\re{bphi}) along each ray
$x=vt+x_0$ with   $|v|\le 1$, then  it holds uniformly in
$x\in\R^n$ owing to (\ref{conp}). We have by (\ref{frepe}),
\be\la{freper}
\Phi(vt+x_0,t)=\fr 1{(2\pi)^n}\sum\limits_{\pm}~~
\int\limits_{\R^n} e^{-i(kv\mp\om)t-ikx_0}
a^\pm(\om)
\hat\Psi( k)~d k.
\ee
This is a sum of oscillatory integrals with the phase
functions $\phi_\pm(k)=kv\pm\om(k)$. Each function
 has two stationary points,
solutions to the equation $v=\pm\na\om(k)$
if $|v|<1$, and has no if $|v|\ge 1$.
The phase  functions are nondegenerate, i.e.
\be\la{Hess}
{\rm det}\left(\ds\frac{\pa^2\phi_\pm(k)}{\pa k_i\pa k_j}\right)_{i,j=1}^n
\ne 0, \,\,\,k\in\R^n.
\ee
At last, $\hat\Psi( k)$ is smooth and decay rapidly at infinity.
Therefore, $\Phi(vt+x_0,t)={\cal O}(t^{-n/2})$
according  to the standard method of
stationary phase, \ci{F}.

{\it Step 2} According to
(\ref{conp})  and
(\ref{bphi}), Eqn (\ref{100rq})
 implies that
\be\la{er}
E|r_t^j|^2\le Ct^{-n}\int\limits_{|x|\le t+{\ov r}}
\chi_r^j(x^n)\Vert q_0(x-y)\Vert ~dx dy
= Ct^{-n}\int\limits_{|x|\le t+{\ov r}}\!\!
\chi_r^j(x^n)dx~
\int\limits_{\R^n}\!\!
\Vert q_0(z)\Vert dz,
\ee
where $\Vert q_0(z)\Vert $ stands for  the norm of a matrix
$\left(q_0^{ij}(z)\right)$.
Therefore,   (\ref{106})  follows as
$\Vert q_0(\cdot)\Vert \in L^1(\R^n)$ by
(\ref{4.7}).
\hfill$\Box$


\setcounter{equation}{0}
\section{Convergence of characteristic functionals}
In this section we complete the proof of  Proposition \ref{l2.2}.
We use a version of the CLT
developed by Ibragimov and Linnik.
If  ${\cal Q}_{\infty}(\Psi,\Psi)=0$,
 Proposition \ref{l2.2} is obvious.
Thus, we may assume that for a given $\Psi\in{\cal D}$,
\be\la{5.*}
{\cal Q}_{\infty}(\Psi,\Psi)\not=0.
\ee
Choose  $0<\de<1$ and
\be\la{rN}
\rho_t\sim t^{1-\delta},
~~~d_t\sim\fr t{\ln t},~~~~\,\,\,t\to\infty.
\ee
\begin{lemma}\la{r}
The following limit holds true:
\be\la{7.15'}
N_t\Bigl(
\varphi(\rho_t)+\Bigl(
\frac{\rho_t}{t}\Bigr)^{1/2}\Bigr)
+
N_t^2\Bigl(
\varphi^{1/2}(\rho_t)+\frac{\rho_t}{t}\Bigr)
\to 0 ,\quad t\to\infty.
\ee
\end{lemma}
{\bf Proof}.
Function $\varphi(r)$ is nonincreasing, hence  by
(\ref{1.12}),
\be\la{1111}
r^{n}\varphi^{1/2}(r)=
n
\int\limits_0^r
s^{n-1}\varphi^{1/2}(r)\,ds\le
n
\int\limits_0^r
s^{n-1}\varphi^{1/2}(s) \,ds\le C\ov\varphi<\infty.
\ee
Then Eqn (\ref{7.15'}) follows as
(\ref{rN}) and (\ref{razl}) imply that $N_t\sim\ln t$.
\hfill$\Box$\\

By the triangle inequality,
\beqn
|\hat\mu_t(\Psi) -\hat\mu_{\infty}(\Psi)|&\le&
|E\exp\{i \langle U_0(t)Y_0,\Psi\rangle  \}-
E\exp\{i{{\sum}}_t r_t^j\}|
\nonumber\\
&&+|\exp\{-\frac{1}{2}{\sum}_t E|r_t^j|^2\} -
\exp\{-\frac{1}{2} {\cal Q}_{\infty}(\Psi, \Psi)\}|
\nonumber\\
&&
+ |E \exp\{i{\sum}_t r_t^j\} -
\exp\{-\frac{1}{2}{\sum}_t E|r_t^j|^2\}|\nonumber\\
&\equiv& I_1+I_2+I_3, \la{4.99}
\eeqn
where the sum ${\sum}_t$ stands for $\sum\limits_{j=-N_t}^{N_t}$.
We are going to   show  that all summands
$I_1$, $I_2$, $I_3$  tend to zero
 as  $t\to\infty$.\\
{\it Step (i)}
Eqn (\ref{razl}) implies
\be\la{101}
 I_1=|E\exp\{i{\sum}_t r^j_t \}
\ds{(\exp\{i{\sum}_t c^j_t\}-1)|\le
 {\sum}_t E|c^j_t|\le
{\sum}_t(E|c^j_t|^2)^{1/2}}.
\ee
>From (\ref{101}), (\ref{106})  and (\ref{7.15'}) we obtain that
\be\la{103}
I_1\le C N_t(\rho_t/t)^{1/2}\to 0,~~t\to \infty.
\ee
{\it Step (ii)}
By the triangle inequality,
\beqn
I_2&\le& \frac{1}{2}
|{\sum}_t E|r_t^j|^2-
 {\cal Q}_{\infty}(\Psi, \Psi) |
\le
 \frac{1}{2}\,
|{\cal Q}_{t}(\Psi, \Psi)-{\cal Q}_{\infty}(\Psi, \Psi)|
\nonumber\\
&&+ \frac{1}{2}\, |E\Bigl({\sum}_t r_t^j\Bigr)^2
-{\sum}_tE|r_t^j|^2| +
 \frac{1}{2}\, |E\Bigl({\sum}_t r_t^j\Bigr)^2
-{\cal Q}_{t}(\Psi, \Psi)|\nonumber\\
&\equiv& I_{21} +I_{22}+I_{23}\la{104},
\eeqn
where ${\cal Q}_{t}$ is  a quadratic form with
the integral kernel $\Big(Q_t^{ij}(x,y)\Big)$.
Eqn (\ref{4.4}) implies that
 $I_{21}\to 0$.
As to  $I_{22}$,  we first have that
\be\la{i22}
I_{22}\le \sum\limits_{j< l}
 E|r_t^j r_t^l|.
\ee
The next lemma   is a corollary of  \ci[Lemma 17.2.3]{IL}.
\begin {lemma}\la{il}
 Let $ \xi$ be a complex random variable
 measurable with respect to
$\sigma$-algebra $\sigma({\cal A})$,
  $\eta$  with respect to
$\sigma$-algebra $\sigma({\cal B})$,
and the distance {\rm dist}$({\cal A}, {\cal B})\ge r>0$. \\
i) Let $(E|\xi|^2)^{1/2}\le a$, $(E|\eta|^2)^{1/2}\le b$. Then
$$
|E\xi\eta-E\xi E\eta|\le
C ab~
\varphi^{1/2}(r).
$$
ii) Let $|\xi|\le a$, $|\eta|\le b$ a.s. Then
$$
|E\xi\eta-E\xi E\eta|\le Cab~
\varphi(r).
$$
\end{lemma}
\bigskip

We apply Lemma \re{il} to deduce that
$I_{22}\to 0$ as $t\to\infty$.
Note  that
$r_t^j=
\langle Y_0(x),\chi_r^j(x^n)({\cal R}_t*\Psi)\rangle $ is measurable
with respect to the $\sigma$-algebra  $\sigma(R_t^j)$.
The distance
between the different rooms $R_t^j$
is greater or equal to
$\rho_t$ according to
 (\ref{rom}).
Then (\ref{i22}) and {\bf S1}, {\bf S3} imply,
together with
Lemma \ref{il} i), that
\be\la{i222}
I_{22}\le
C N_t^2\varphi^{1/2}(\rho_t),
\ee
which goes to $0$ as $t\to\infty$ because of
(\ref{106}) and (\ref{7.15'}).
Finally, it remains to check
 that $I_{23}\to 0$,
$t\to\infty$. By the
Cauchy - Schwartz inequality,
\beqn
I_{23}&\le&
 |E\Bigl({\sum}_t r_t^j\Bigr)^2
- E\Bigl({\sum}_t r_t^j +
{\sum}_t c_t^j\Bigr)^2 |\nonumber\\
& \le&
C N_t{\sum}_t E |c_t^j|^2  +
C\Bigl(
E({\sum}_t r_t^j)^2\Bigr)^{1/2}
\Bigl(
N_t{\sum}_t E|c_t^j|^2\Bigr)^{1/2}.\la{107}
\eeqn
Then  (\ref{106}), (\ref{i22}) and (\ref{i222})
imply
$$
E({\sum}_t r_t^j)^2\le
{\sum}_tE|r_t^j|^2 +2{\sum}_{j<l}E|r_t^j r_t^l|
\le
CN_td_t/t+C_1N_t\varphi^{1/2}(\rho_t)\le C_2<\infty.
$$
Now
(\ref{106}),  (\ref{107}) and (\ref{7.15'}) yields
\be\la{106'}
I_{23}\le C_1  N_t^2\rho_t/t+C_2 N_t(\rho_t/t)^{1/2} \to 0,~~t\to \infty.
\ee
So,  all terms $I_{21}$, $I_{22}$, $I_{23}$
in  (\ref{104})
tend to zero.
Then
 (\ref{104}) implies that
\be\la{108}
I_2\le
\frac{1}{2}\,
|{\sum}_{t}E|r_t^j|^2-
 {\cal Q}_{\infty}(\Psi, \Psi)|
\to 0,~~t\to\infty.
\ee
{\it Step (iii)}
It remains to verify that
\be\la{110}
I_3=|
E\exp\{i{\sum}_t r_t^j\}
-\exp\{-\fr12E\Big({\sum}_t r_t^j\Big)^2\}| \to 0,~~t\to\infty.
\ee
Using Lemma \ref{il}, ii)
yields:
\beqn
&&|E\exp\{i{\sum}_t r_t^j\}-\prod\limits_{-N_t}^{N_t}
E\exp\{i r_t^j\}|
\nonumber\\
&\le&
|E\exp\{ir_t^{-N_t}\}\exp\{i\sum\limits_{-N_t+1}^{N_t} r_t^j\}  -
 E\exp\{ir_t^{-N_t}\}E\exp\{i\sum\limits_{-N_t+1}^{N_t} r_t^j\} |
\nonumber\\
&&+
|E\exp\{ir_t^{-N_t}\}E\exp\{i\sum\limits_{-N_t+1}^{N_t} r_t^j\}
-\prod\limits_{-N_t}^{N_t}
E\exp\{i r_t^j\}|
\nonumber\\
&\le& C\varphi(\rho_t)+
|E\exp\{i\sum\limits_{-N_t+1}^{N_t} r_t^j\}
-\prod\limits_{-N_t+1}^{N_t}
E\exp\{i r_t^j\}|.\nonumber
\eeqn
We then apply Lemma \ref{il}, ii) recursively
and get, according to Lemma \ref{r},
\be\la{7.24'}
|E\exp\{i{\sum}_{t} r_t^j\}-\prod\limits_{-N_t}^{N_t}
E\exp\{i r_t^j\}|
\le
C N_t\varphi(\rho_t)\to 0,\quad t\to\infty.
\ee
It remains to check that
\be\la{110'}
|\prod\limits_{-N_t}^{N_t} E\exp\{ir_t^j\}
-\exp\{-\fr12{\sum}_{t} E|r_t^j|^2\}| \to 0,~~t\to\infty.
\ee
According to the standard statement of the CLT
(see, e.g. \ci[Thm 4.7]{P}),
it suffices to verify the  Lindeberg condition:
$\forall\ve>0$
\be\la{Lind}
\frac{1}{\sigma_t}
{\sum}_t
 E_{\ve\sqrt{\sigma_t}}
|r_t^j|^2 \to 0,~~t\to\infty.
\ee
Here
$
\sigma_t\equiv {\sum}_t
E |r^j_t|^2,$
and $E_\de f\equiv E X_\de f$,
where $X_\de$ is the indicator of the
event $|f|>\de^2.$
Note that
(\ref{108})
and (\re{5.*}) imply  that
$$
\sigma_t \to
{\cal Q}_{\infty}(\Psi, \Psi)\not= 0,~~t\to\infty.
$$
Hence it remains to verify that $\forall\ve>0$
\be\la{linc}
{\sum}_t
E_{\ve}
|r_t^j|^2 \to 0,~~t\to\infty.
\ee
We check Eqn (\ref{linc}) in Section 9. This will complete the proof of
Proposition \ref{l2.2}.
\hfill$\Box$


\setcounter{equation}{0}
\section{The Lindeberg condition}
The proof of (\ref{linc}) can be reduced to the case when
for some $\La\ge 0$ we have, almost sure that
 \be\la{ass1}
  |u_0(x)|+|v_0(x)|\leq \La<\infty, ~~~x\in\R^n.
  \ee
Then the proof of (\ref{linc}) is reduced  to
the convergence
\be\la{111}
{\sum}_t E|r_t^j|^4 \to 0,~~t\to\infty
\ee
by using Chebyshev's inequality.
The general case can be covered by standard cutoff arguments
by
taking into account that
bound (\ref{106}) for $E|r^j_t|^2$
depends only on $e_0$ and $\varphi$.
 The last fact is obvious  from
(\ref{er}) and (\ref{qp}) with $p=1$ and $\ga=0$.

We deduce (\ref{111}) from
\begin{theorem}  \la{p5.1}
Let the conditions  of Theorem B  hold
and assume that (\ref{ass1}) is fulfilled.
Then
for any $\Psi\in {\cal D}$
there exists a constant $C(\Psi)$
such that
\be\la{112}
E|r_t^j|^4\le
C(\Psi) \La^4d_t^2/t^2,   ~~t>1.
\ee
\end{theorem}
{\bf Proof.} {\it Step 1}~
Given four points $x_1,x_2,x_3,x_4\in\R^n$, set:
$$
M_0^{(4)}(x_1,...,x_4)=E\left(Y_0(x_1)\otimes...\otimes Y_0(x_4)\right).
$$
Then, similarly to (\ref{100rq}),
Eqns (\ref{ass1}) and (\ref{100}) imply by
the Fubini Theorem that
\be\la{11.2}
E|r_t^j|^4=
\langle \chi_r^j(x_1^n) \ldots \chi_r^j(x_4^n)M_0^{(4)}(x_1,\dots,x_4),
\Phi(x_1,t)\otimes\dots\otimes\Phi(x_4,t)\rangle .
\ee
Let us analyse the domain of the integration ${(\R^n)^4}$ in
the RHS of (\ref{11.2}).
We partition  ${(\R^n)^4}$ into three parts $W_2$,  $W_3$  and $W_4$:
\be\la{Wi}
{(\R^n)^4}=
\bigcup\limits_{i=2}^4 W_i,\quad
W_i=\{\bar x=(x_1,x_2,x_3,x_4)\in {(\R^n)^4}:
|x_1-x_i|=\max\limits_{p=2,3,4}|x_1-x_p|\}.
\ee
Furthermore,  given
 $\bar x=(x_1,x_2,x_3,x_4)\in W_i$,
  divide $\R^n$ into three parts
$S_j$, $j=1,2,3$: $\R^n=S_1\cup S_2\cup S_3$,
by two
hyperplanes orthogonal to
the segment $[x_1,x_i]$ and  partitioning it into three equal segments,
where  $x_1\in S_1$ and $x_i\in S_3$.
Denote by
$x_p$, $x_q$  the  two remaining points
with $p,q\ne 1,i$.
Set:
 ${\cal A}_i=\{\bar x\in W_i:~
 x_p\in S_1,  x_q\in S_3 \}$,
 ${\cal B}_i=\{\bar x\in W_i:~
 x_p, x_q\not\in S_1 \}$ and
 ${\cal C}_i=\{\bar x\in W_i:~
 x_p, x_q\not\in S_3 \}$, $i=2,3,4$.
 Then
$W_i={\cal A}_i\cup {\cal B}_i\cup {\cal C}_i$.
Define the function ${\rm m}^{(4)}_0(\bar x)$, $\bar x\in{(\R^n)^4},$
in the following way:
\beqn\la{M}
{\rm m}^{(4)}_0(\bar x)\Bigr|_{W_i}=
\left\{
\ba{ll}
M_0^{(4)}(\bar x)-
q_0(x_1-x_p)\otimes
 q_0(x_i-x_q),\quad \bar x\in{\cal A}_i,\\
M_0^{(4)}(\bar x),\quad \bar x\in{\cal B}_i\cup {\cal C}_i.
\ea
\right.
\eeqn
This determines ${\rm m}^{(4)}_0(\bar x)$
correctly  for almost all quadruples  $\bar x$.
Note that
\beqn
\!&&\!\!\!\langle \chi_r^j(x_1^n) \ldots \chi_r^j(x_4^n)
q_0(x_1-x_p)\otimes q_0(x_i-x_q),
\Phi(x_1,t)\otimes\dots\otimes\Phi(x_4,t)\rangle
\nonumber\\
\!&&\!\!\!\!\!\!\!\!\!=\langle \chi_r^j(x_1^n)\chi_r^j(x_p^n) q_0(x_1-x_p),
\Phi(x_1,t)\otimes\Phi(x_p,t)\rangle~
\langle \chi_r^j(x_i^n)\chi_r^j(x_q^n) q_0(x_i-x_q),
\Phi(x_i,t)\otimes\Phi(x_q,t)\rangle .
\nonumber
\eeqn
Each factor here is bounded by
$ C(\Psi)~d_t/t$.
Similarly to  (\ref{106}),
this can be deduced  from an expression of type (\ref{100rq})
for the factors.
Therefore, the proof of (\ref{112}) reduces to the proof
of the  bound
\be\la{st1}
I_t:=|\langle \chi_r^j(x_1^n)\ldots\chi_r^j(x_4^n){\rm
m}^{(4)}_0(x_1,\dots,x_4),
\Phi(x_1,t)\otimes\dots\otimes\Phi(x_4,t)\rangle |\le
C(\Psi) \La^4d_t^2/t^2,\quad t > 1.
\ee
{\it Step 2}~
Similarly to (\re{er}),
 Eqn (\ref{bphi}) implies,
\be\la{500}
I_t\le
 C(\Psi)~t^{-2n}\int\limits_{ (B_t^{\ov r})^4 }
\chi_r^j(x_1^n) \ldots \chi_r^j(x_4^n)
 |{\rm m}^{(4)}_0(x_1,\dots,x_4)| dx_1~dx_2~dx_3~ dx_4,
\ee
where $B_t^{\ov r}$ is the ball
$\{x\in\R^n:~|x|\le t+{\ov r}\}$.
Let us estimate ${\rm m}^{(4)}_0$
using Lemma  \ref{il}, ii).
\begin{lemma}\la{l11.1}
For each $i=2,3,4$ and almost all  $\ov x\in W_i$ the
following bound holds
\be\la{115}
 |{\rm m}^{(4)}_0(x_1,\dots,x_4)|
\le
C \La^4\varphi(|x_1-x_i|/3).
\ee
\end{lemma}
{\bf Proof.}
For $\bar x\in {\cal A}_i$ we apply
 Lemma  \ref{il}, ii)
to $\C^2\otimes\C^2\equiv \R^4\otimes\R^4$-valued random variables
$\xi=Y_0(x_1)\otimes Y_0(x_p)$ and
 $\eta=Y_0(x_i)\otimes Y_0(x_q)$. Then
(\ref{ass1}) implies the bound  for almost all $\bar x\in {\cal A}_i$
\be\la{M1}
|{\rm m}^{(4)}_0(\bar x)|\le
C \La^4 \varphi(|x_1-x_i|/3).
\ee
For  $\bar x\in {\cal B}_i$,
we apply
 Lemma  \ref{il}, ii)
to
$\xi=Y_0(x_1)$ and
 $\eta= Y_0(x_{p})\otimes Y_0(x_{q})\otimes Y_0(x_{i})$. Then
  {\bf S0}  implies a similar bound for
almost all $\bar x\in {\cal B}_i$,
\be\la{113}
|{\rm m}^{(4)}_0(\bar x)|=
| M_0^{(4)}(\bar x) -
EY_0(x_1)\otimes
E\Bigl( Y_0(x_{p})\otimes Y_0(x_{q})\otimes Y_0(x_{i})\Bigr)|
\le C \La^4 \varphi(|x_1-x_i|/3),
\ee
and the same for
almost all
$\bar x\in {\cal C}_i$. \hfill$\Box$
\medskip\\
{\it Step 3}~It remains to prove the following bounds
for each $i=2,3,4$:
\be\la{vit}
V_i(t):=\int\limits_{ (B_t^{\ov r})^4 }
\chi_r^j(x_1^n) \ldots \chi_r^j(x_4^n)
 X_i(\ov x)   \varphi(|x_1-x_i|/3)
dx_1~dx_2~dx_3~ dx_4
\le Cd_t^2 t^{2n-2},
\ee
where $X_i$ is an indicator of the set $W_i$.
In fact, this integral does not depend on $i$,  hence
set $i=2$ in the integrand:
\be\la{500s}
V_i(t)
\le
 C\int\limits_{ (B_t^{\ov r})^2 }
\chi_r^j(x_1^n) \varphi(|x_1-x_2|/3)
\left[\int\limits_{ B_t^{\ov r} }\chi_r^j(x_3^n)
\left(\int\limits_{ B_t^{\ov r} }
X_2(\ov x)~dx_4\right)~dx_3 \right]
 dx_1~dx_2.
\ee
Now a key observation is that the inner integral  in $dx_4$
is ${\cal O}(|x_1-x_2|^n)$ as $X_2(\ov x)=0$ for
$|x_4-x_1| > |x_1-x_2|$. This implies
\be\la{500s4}
V_i(t)
\le
 C{\ov r}^4\int\limits_{ B_t^{\ov r}  }\chi_r^j(x_1^n)
\left(\int\limits_{ B_t^{\ov r}  }
 \varphi(|x_1-x_2|/3)
|x_1-x_2|^n~ dx_2\right)~dx_1
 \int\limits_{ B_t^{\ov r}  }\chi_r^j(x_3^n)
~dx_3.
\ee
The inner integral in $dx_2$ is bounded as
\beqn\la{qphi4}
&&\int\limits_{ B_t^{\ov r}}
 \varphi(|x_1-x_2|/3)
|x_1-x_2|^n~ dx_2
\le C(n)\int_0^{2(t+{\ov r})}r^{2n-1} \varphi( r/3)~dr\nonumber\\
&&\le
C_1(n) \sup\limits_{r\in [0,2(t+{\ov r})]} ~r^n\varphi^{1/2}( r/3)
\int_0^{2(t+{\ov r})}r^{n-1} \varphi^{1/2}(  r/3)~dr,
\eeqn
where the `$\sup$' and the last integral are bounded by
(\re{1111}) and (\re{1.12}), respectively. Therefore,  (\re{vit})
follows from (\re{500s4}).
This completes the proof of
Theorem \ref{p5.1}.
\hfill$\Box$\\
~\\
{\bf Proof of convergence (\ref{111}).}
As $d_t\le h\sim t/N_t$,
bound  (\ref{112}) implies,
$$
{\sum}_t
E|r_t^j|^4  \le
\fr{C\La^4d_t^2}
{t^2}N_t
\le
\fr{C_1\La^4}
{N_t}
\to 0,\quad N_t\to\infty.
$$
\hfill$\Box$


\setcounter{equation}{0}
\section{
The scattering theory for infinite energy solutions}

In this section we develop a version of the
 scattering theory  to deduce
Theorem A
from  Theorem B.
The main step is to establish
an asymptotics of type  (\re{dsti}) for adjoint groups
by using  results of Vainberg \ci{V74}.

Consider  operators $U'(t)$,   $U'_0(t)$ in the complex space
$ H =L^2(\R^n)\oplus H^1(\R^n)$ (see (\ref{1.20})).
The
energy
conservation for the KGE
implies  the following  corollary:
\begin{cor}\la{co7.1}
There exists a constant $C>0$ such that
$\forall\Psi\in H$:
\be\la{6.4}
\Vert U'_0(t)\Psi\Vert_{ H} \le C
\Vert \Psi\Vert_{ H},\,\,\,\,\,
\Vert U'(t)\Psi\Vert_{ H} \le C
\Vert \Psi\Vert_{ H},\,\,\,\,t\in\R.
\ee
\end{cor}
Lemma  \re{l6.1} below
develops earlier results \ci[Thms 3,4,5]{V74}.
Consider a family of finite  seminorms in $ H$
$$
\Vert\Psi\Vert_{(R)}^2= \int\limits_{|x|\le R}
(|\Psi_0(x)|^2+|\Psi_1(x)|^2+|\nabla\Psi_1(x)|^2)\,dx,~~~~~ R>0.
$$

Denote by
$ H_{(R)}$ the subspace
of  functions from $ H$ with a support in the ball
$B_R$.

\begin{definition}
 $H_c$ denotes the space $\cup_{R>0} H_{(R)}$
endowed with the following convergence:
a  sequence $\Psi_n$ converges to $\Psi$ in $H_c$ iff
$\exists R>0$ such that  all  $\Psi_n\in H_{(R)}$, and
$\Psi_n$ converge to $\Psi$ in the norm $\Vert\cdot\Vert_{(R)}$.
\end{definition}

 Below, we
speak of  continuity of  maps from
$H_c$  in the sense of  sequential
continuity.
Given $t\ge 0$, denote
\be\la{eps}
\ve(t)= \left\{\ba{ll}
(t+1)^{-3/2},&n\ge 3,\\
(t+1)^{-1}\ln^{-2}(t+2),&n=2.
\ea
\right.
\ee
\begin{lemma} \la{l6.1}
Let Assumptions {\bf E1} -- {\bf E3} hold, and $n\ge 2$.
Then
 for any $R,R_0>0$ there exists a constant $C=C(R,R_0)$
such that for $\Psi\in  H_{(R)}$
\be\la{6.5}
\Vert U'(t)\Psi\Vert_{(R_0)} \le C
\ve(t)
\Vert \Psi\Vert_{(R)},\,\,t\ge 0.
\ee
\end{lemma}

This lemma has been proved in \ci{diss}
by using  Conditions  {\bf E1}-{\bf E3} and
a  method  developed in \ci{V89}.
For the proof, the contour of integration in the $k$-plane
from \ci{V74} had to be curved loragithmically at infinity
as in  \ci{V89},
but should not be chosen
 parallel to the real axis.
\medskip

The main result of this section is Theorem \re{t6.1} below.
Given $t\ge 0$, set
\be\la{6.7}
\ve_1(t)= \left\{\ba{ll}
(t+1)^{-1/2},&n\ge 3,\\
\ln^{-1}(t+2),&n=2.
\ea
\right.
\ee
\begin{theorem}\la{t6.1}
Let Assumptions {\bf E1}-{\bf E3} and {\bf S0}-{\bf S3}  hold, and $n\ge 2$.
Then there exist linear continuous operators
$  W,r(t):~{ H_c}\to  H$
such that
for $\Psi\in{ H_c}$
\be\la{dst}
U^\pr(t)\Psi=U^\pr_0(t)W\Psi+r(t)\Psi, \,\,\,t\ge 0,
\ee
and the following  bounds hold $\forall R>0$ and
 $\Psi\in  H_{(R)}$:
\beqn
\Vert r(t)\Psi\Vert_ H&\le
C(R)\ve_1(t)     \Vert\Psi\Vert_{(R)},&\,\,t\ge 0,
\la{6.9}\\
E|\langle Y_0,r(t)\Psi\rangle |^2&\le
C(R)\ve_1^2(t)   \Vert\Psi\Vert_{(R)}^2,&\,\,
 t\ge 0.
\la{6.6}
\eeqn
\end{theorem}
{\bf Proof.} We apply the standard Cook method:
see, e.g.,  \ci[Thm XI.4]{RS3}.
Fix $\Psi\in  H_{(R)}$
and
define $  W\Psi$,  formally, as
$$
  W\Psi=\lim_{t\to\infty}U'_0(-t)U'(t)\Psi  =
\Psi+
\int\limits_{0}^{\infty}\frac{d}{dt}
U'_0(-t)U'(t)\Psi  \,dt.
$$
We have to prove
the convergence of the integral in  norm in  space
$ H$.
 First, observe that
$$
\frac{d}{dt} U'_0(t)\Psi={\cal A}'_0 U'_0(t)\Psi,\,\,\,\,
~~~~
\frac{d}{dt} U'(t)\Psi={\cal A}' U'(t)\Psi,
$$
where
${\cal A}'_0$ and  ${\cal A}'$
are the generators to  groups $U'_0(t)$,  $U'(t)$,
respectively.
Similarly to (\ref{A0'}), we have
\be\la{A'}
{\cal A}'=\left( \begin{array}{cc} 0 & A \\
1 & 0 \end{array}\right),
\ee
where
 $A=
\sum\limits_{j=1}^{n}
(\partial_j-iA_k)^2-m^2$.
Therefore,
\be\la{6.10}
\frac{d}{dt} U'_0(-t) U'_1(t)\Psi
=U'_0(-t) ({\cal A}'-{\cal A}'_0) U'(t)\Psi.
\ee
Now (\ref{A'}) and (\ref{A0'}) imply
$$
{\cal A}'-{\cal A}'_0 =
\left( \begin{array}{cc} 0 & L  \\
0 & 0 \end{array}\right).
$$
Furthermore, {\bf E2} implies that
 $L=\sum\limits_{j=1}^{n}(\partial_j-iA_j)^2-\Delta$
is a
first order partial differential operator
with the coefficients  vanishing for $|x|\ge R_0$.
Thus,
 (\ref{6.4}) and (\ref{6.5}) imply that
\beqn
\Vert U'_0(-t) ({\cal A}'-{\cal A}'_0) U'(t)\Psi
\Vert_ H&\le&
C~\Vert({\cal A}'-{\cal A}'_0) U'(t)\Psi\Vert_ H
=
C~\Vert\Big(({\cal A}'-{\cal A}'_0) U'(t)\Psi\Big)^0
\Vert_{L^2(B_{R_0})}
\nonumber\\
&\le&
 C_1~\Vert\Big( U'(t)\Psi\Big)^1\Vert_{H^1(B_{R_0})}
\le
C(R)\ve(t)  \Vert\Psi\Vert_{(R)},
~  t\ge 0.\la{6.11}
\eeqn
Hence  (\ref{6.10})  implies
\be\la{6.12}
\int\limits_{s}^{\infty}
\Vert\frac{d}{dt} U'_0(-t) U'(t)\Psi\Vert_ H\,dt
\le C(R)  \ve_1(s) \Vert\Psi\Vert_{(R)},\,\,\,s\ge 0.
\ee
Therefore, (\re{dst}) and (\re{6.9}) follow by (\re{6.4}).
It remains to prove (\re{6.6}).
First, similarly to (\re{100rq}),
\be\la{6.14s}
E\langle Y_0,r(t)\Psi\rangle ^2
=
\langle q_0(x-y),
 r(t)\Psi(x)\otimes
 {r(t)\Psi(y)}~
\rangle
\ee
Therefore, the Shur Lemma implies (similarly to (\re{er}))
\be\la{6.14}
E\langle Y_0, r(t)\Psi\rangle ^2
\le
\Vert q_0\Vert_{L^1}\,\,
\Vert  r(t)\Psi \Vert_{L^2}\,\,
\Vert  r(t)\Psi \Vert_{L^2},
\ee
where the norms $\Vert\cdot\Vert_{L^p}$ have an obvious meaning.
Finally,
 (\ref{6.9}) implies
for
$\Psi\in  H_{(R)}$
\be\la{6.15}
\Vert  r(t)\Psi \Vert_{L^2}
\le C
\Vert r(t)\Psi\Vert_ H\le
C(R)\ve_1(t)     \Vert\Psi\Vert_{(R)},
\ee
Therefore, (\ref{6.6}) follows from (\ref{6.14})
since $\Vert q_0\Vert_{L^1}<\infty$ by
(\ref{4.7}).
\hfill$\Box$

\setcounter{equation}{0}
\section{Convergence to equilibrium for variable coefficients}
The assertion of Theorem A follows from two
propositions below:
\begin{pro}\la{p9.1}
The family of the measures $\{\mu_t, t\in\R\}$,
is weakly compact in ${\cal H}^{-\ve}$, $\forall\ve>0$.
\end{pro}
\begin{pro}\la{p9.2}
For any $\Psi\in{\cal D}$
\be\la{2.6'}
 \hat \mu_t(\Psi )\equiv\int \exp(i\langle Y,\Psi\rangle )~\mu_t(dY)
 \rightarrow
\exp\{-\fr12{\cal Q}_\infty (W\Psi, {W\Psi}~)\},
 \,\,\,t\to\infty.
 \ee
\end{pro}
We deduce these propositions from Propositions
\re{l2.1} and \re{l2.2}, respectively, with the help of Theorem
\ref{t6.1}.
\medskip\\
{\bf Proof of Proposition \ref{p9.1}} Similarly to Proposition
\re{l2.1}, Proposition \re{p9.1} follows from the bounds
\be\la{7.0}
\sup\limits_{t\ge 0}E\Vert U(t)Y_0\Vert_{R}<\infty,\,\,\,\,\,R>0.
\ee
For the proof, write the solution to (\ref{1.3})
in the form
\be\la{7.3}
u(x,t)=v(x,t)+w(x,t).
\ee
Here
$v(x,t)$ is the solution to  (\ref{2.1}), and
$w(x,t)$ is the solution to the following Cauchy problem
\beqn\la{7.1}
\left\{
\ba{rcl}
 \ddot w(x,t) &=&\,\,\,
\sum_{k=1}^{n}(\pa_k - iA_{k}(x))^2 w(x,t)
 - m^2\, w(x,t)\\
~\\
&&\!\!-\sum_{k=1}^{n}2iA_{k}(x)\pa_k v(x,t)-
\sum_{k=1}^{n}(i\pa_k A_{k}(x)+A^2_{k}(x))v(x,t),\\
~\\
 w|_{t=0}&=&0,~~\dot w|_{t=0} = 0,~~ x\in\R^n.
\ea
\right.
 \eeqn
Then (\ref{7.3}) implies
\be\la{7.4}
E\Vert U(t)Y_0\Vert_{R}\le
E\Vert U_0(t)Y_0\Vert_{R} +
E\Vert(w(\cdot,t),\dot w(\cdot,t))\Vert_{R}.
\ee
By Proposition \ref{p2.1} we have
\be\la{7.5}
\sup\limits_{t\ge 0}E\Vert U_0(t)Y_0\Vert_{R}<\infty.
\ee
It remains to estimate the second term
in the right hand side of (\ref{7.4}).
The Duhamel representation for the solution to
(\ref{7.1}) gives
\be\la{Duh}(w,\dot w) =\int\limits_0^t
U(t-s)(0,\psi(\cdot,s))\,ds,\ee
where $\psi(x,s)=-2i\sum\limits_{k=1}^{n}A_{k}(x)
\pa_k v(x,s)-
\sum\limits_{k=1}^{n}(i\pa_k A_{k}(x)+A^2_{k}(x))v(x,s)$.
Assumption {\bf E2} implies that
$\supp \psi(\cdot,s)\subset B_{R_0}$.
Moreover,
\be\la{mo}
\Vert(0,\psi(\cdot,s))\Vert_{R_0}\le C
\Vert v(\cdot,s)\Vert_{H^1(B_{R_0})}\le C\Vert U_0(s)Y_0\Vert_{R_0}.
\ee
The
 decay estimates of type (\ref{6.5}) hold for  the group $U(t)$,
as well as for  $U'(t)$,
as  both groups  correspond to the same equation
by Lemma \re{ldu}.
Hence, we have  from  (\ref{mo}),
\be\la{7.6}
\Vert U(t-s)(0,\psi(\cdot,s))\Vert_{R}\le
C(R)\ve(t-s)\Vert(0,\psi(\cdot,s))\Vert_{R_0}
\le
C_1(R)\ve(t-s)\Vert U_0(s)Y_0\Vert_{R_0},
\ee
where $\ve(\cdot)$ is defined in (\re{eps}).
Therefore,  (\ref{Duh}) and (\ref{7.5})  imply
\be\la{7.7}
E\Vert(w(\cdot,t),\dot w(\cdot,t))\Vert_{R}
\le C(R)\int\limits_0^t
\ve (t-s)E\Vert U_0(s)Y_0\Vert_{R_0}\,ds
\le C_2(R)<\infty,\,\,\,\,\,t\ge 0.
\ee
Then   (\ref{7.5}) and (\ref{7.4})
imply
 (\ref{7.0}).
\hfill$\Box$
 \medskip\\
{\bf Proof of Proposition \ref{p9.2}}
(\ref{dst}) and
 (\ref{6.6}) imply by Cauchy-Schwartz,
$$
\ba{l}
|E\exp{i\langle U(t)Y_0,\Psi\rangle }-E\exp{i\langle Y_0,U'_0(t)  W\Psi\rangle
}|\le
E|\langle Y_0,r(t)\Psi\rangle |\\
~\\
\le (E|\langle Y_0,r(t)\Psi\rangle |^2)^{1/2}
\to 0,~~t\to\infty.
\ea
$$
It remains to prove that
\be\la{7.10}
E \exp{i\langle Y_0,U'_0(t)  W\Psi\rangle }  \to
\exp\{-\fr12{\cal Q}_\infty(  W\Psi,  {W\Psi}~)\},
~~t\to\infty.
\ee
This does not follow directly  from
Proposition \ref{l2.2} since generally,
$W\Psi\not\in{\cal D}$. We approximate $W\Psi$
by  functions from ${\cal D}$.
$W\Psi\in H$, and  ${\cal D}$ is dense in $H$.
Hence, for any $\eps>0$
there exists $\Phi\in{\cal D}$  such that
\be\la{7.12}
\Vert    W\Psi-\Phi  \Vert_ H
\le\eps.
\ee
Therefore, we can derive (\ref{7.10})
by the triangle
inequality
\beqn
&&|E \exp{i\langle Y_0,U'_0(t)  W\Psi\rangle }-
\exp\{-\fr12{\cal Q}_\infty(W\Psi, W {\Psi}~)\}|\nonumber\\
&&\le
|E \exp{i\langle Y_0,U'_0(t)  W\Psi\rangle }-
E\exp{i\langle Y_0,U'_0(t)\Phi\rangle }|\nonumber\\
&&
+E|\exp{i\langle U_0(t)Y_0,\Phi\rangle }-
\exp\{-\fr12{\cal Q}_\infty(\Phi, {\Phi}~)\}|
\nonumber\\
&&+|\exp\{-\fr12{\cal Q}_\infty(\Phi, {\Phi}~)\}-
\exp\{-\fr12{\cal Q}_\infty(  W\Psi,  {W\Psi}~)\}|.
\la{7.13}
\eeqn
Applying
Cauchy-Schwartz, we get,
similarly to (\ref{6.14s})-(\ref{6.15}), that
$$E|\langle Y_0,U'_0(t)(  W\Psi-\Phi)\rangle |\le
(E|\langle Y_0,U'_0(t)(  W\Psi-\Phi)\rangle |^2)^{1/2}\le
C\Vert U'_0(t)(W\Psi-\Phi)\Vert_H.$$
Hence,  (\ref{6.4}) and (\ref{7.12}) imply
\be\la{7.11}
E|\langle Y_0,U'_0(t)(  W\Psi-\Phi)\rangle |\le
C\eps,\,\,\,t\ge 0.
\ee
Now we can estimate each term in the right hand side of (\ref{7.13}).
The first
term
is ${\cal O}(\eps)$
 uniformly in $t>0$ by (\ref{7.11}).
The second term
converges to zero as $t\to\infty$ by Proposition \ref{l2.2}
since  $\Phi\in {\cal D}$.
Finally, the third term
is ${\cal O}(\eps)$
owing to  (\ref{7.12})  and  the continuity
of the quadratic form ${\cal Q}_\infty(\Psi, \Psi)$
in $L^2(\R^n)\otimes\C^2$. The continuity follows
from the Shur Lemma since the integral kernels
$q_\infty^{ij}(z)\in L^1(\R^n)\otimes M^2$ by Corollary \re{coro}.
Now the  convergence in (\ref{7.10})  follows
since $\eps>0$ is arbitrary.
\hfill$\Box$

\setcounter{equation}{0}
\section{Appendix A. Fourier transform calculations}

Consider the covariance functions of the solutions
to the system (\ref{2'}). Let  $F:~w\mapsto\hat w$
denote the FT of a tempered distribution
$w\in S'(\R^n)$ (see, e.g. \ci{EKS}).
We  also use this notation for vector-
and matrix-valued functions.

\subsection{Dynamics in the FT space}
In the FT representation, the system (\ref{2'}) becomes
$\dot {\hat Y}(k,t)=\hat{\cal A}_0 (k)\hat Y(k,t)$, hence
\be\la{Frep}
\hat Y(k,t)=\hat{\cal G}_t( k)
\hat Y_0(k),
\,\,\,\,\,\,\hat{\cal G}_t( k)=\exp({\hat{\cal A}_0(k)t}).
\ee
Here we denote
\be\la{hatA}
\hat{\cal A}_0(k)=
\left( \begin{array}{ccc} 0 &~~& 1 \\
~\\
 -|k|^2-m^2 &~~& 0 \end{array}\right),
\,\,\,\,\,\,\,\,\,\,\quad\quad
\hat{\cal G}_t( k)=
\left( \begin{array}{ccc} {\rm cos}~\om t &~~& \ds\fr{\sin \om t}{\om}  \\
~\\
 -\om~{\rm sin}~\om t
&~~&  {\rm cos}~\om t\end{array}\right),
\ee
where $\om=\om( k)=\sqrt{ |k|^2+m^2}$.

\subsection{Covariance matrices in the FT space}
\begin{lemma}\la{lfq}
In the sense of matrix-valued distributions
\be\la{tidtx}
q_t(x-y):=
E \Big(Y(x,t)\otimes{Y(y,t)}~\Big)=F^{-1}_{k\to x-y}
\hat{\cal G}_t( k)\hat q_0( k)\hat{\cal G}'_t( k),
\,\,\,\,t\in\R.
\ee
 \end{lemma}
{\bf Proof}
Translation invariance  (\ref{1.9'}) implies
\be\la{tic}
E\Big(  Y_0(x)\otimes_C{ Y_0(y)}~\Big)=C^+_0(x-y),\,\,\,\,
E\Big(  Y_0(x)\otimes_C\ov{ Y_0(y)}~\Big)=C^-_0(x-y),
\ee
where $\otimes_C$ stands for tensor product of complex vectors.
Therefore,
\beqn\la{tid}
E\Big( \hat Y_0(k)\otimes_C{\hat Y_0(k')}~\Big)=F_{x\to k}F_{y\to k'}
\,\,C^+_0(x-y)=
(2\pi)^n
\de( k+k')\hat C^+_0( k),\nonumber\\
E\Big( \hat Y_0(k)\otimes_C\ov{\hat Y_0(k')}~\Big)=
F_{x\to k}F_{y\to -k'}C^-_0(x-y)=
(2\pi)^n
\de( k-k')\hat C^-_0( k).
\eeqn
Now (\ref{Frep}) and (\ref{hatA}) give
in the matrix notation that
\beqn\la{tidt}
E\Big(\hat Y(k,t)\otimes_C{\hat Y(k',t)}~\Big)=
(2\pi)^n
\de( k+k')\hat{\cal G}_t( k)\hat C^+_0( k)\hat{\cal G}'_t( k),\nonumber\\
E\Big(\hat Y(k,t)\otimes_C\ov {\hat Y(k',t)}~\Big)=
(2\pi)^n
\de( k-k')\hat{\cal G}_t( k)\hat C^-_0( k)\hat{\cal G}'_t( k).
\eeqn
Therefore, by the inverse FT formula we get
\beqn\la{tidty}
E\Big( Y(x,t)\otimes_C{ Y(y,t)}~\Big)= F^{-1}_{k\to x-y}
\hat{\cal G}_t( k)\hat C^+_0( k)\hat{\cal G}'_t( k),\nonumber\\
E\Big( Y(x,t)\otimes_C\ov { Y(y,t)}~\Big)=
F^{-1}_{k\to x-y}\hat{\cal G}_t( k)\hat C^-_0( k)\hat{\cal G}'_t( k).
\eeqn
Then
(\re{tidtx}) follows by linearity.\hfill$\Box$

\setcounter{equation}{0}
\section{Appendix B. Measures
 in Sobolev's spaces}

Here we  formally verify bound (\re{Min}) for $s,\al<-n/2$.
Definition (\re{ws}) implies for
$u\in{ H}^{s,\al}$,
\be\la{wsf}
\Vert u\Vert_{s,\al}^2=\fr 1{(2\pi)^{2n}}
\int\langle x \rangle^{2\al}
\Big[
\int \ds e^{-ix( k-k')}
\langle k\rangle^s\langle k'\rangle^s
 \hat u( k)\ov{\hat u(k')}\,d k dk'
\Big]dx.
\ee
Let $\mu(du)$ be a  translation-invariant measure
in ${ H}^{s,\al}$ with a CF $Q(x,y)=q(x-y)$.
Similarly to  (\re{tid}), (\re{tic}), we get
\be\la{corfuex}
\int
\hat u( k)\ov{\hat u(k')}
\mu(du)=
(2\pi)^n
\de( k-k')~\tr\hat q( k).
\ee
Then,
integrating (\re{wsf}) with respect to the measure
$\mu(du)$, we get the formula
\be\la{wsmu}
\int
\Vert u\Vert_{s,\al}^2
\mu(du)=
\fr 1{(2\pi)^n}
\int\langle x\rangle^{2\al}dx
\int
\langle k\rangle^{2s} ~\tr\hat q( k)\,
d k.
\ee
Applying it to $ \hat q( k)=T$
with   $\al,s<-n/2$
and to
$ \hat q( k)=T(k^2+m^2)^{-1}$
with $1+s$ instead of $s$,
we get (\re{Min}).


   \end{document}